\documentclass[fleqn,usenatbib]{mnras}
\usepackage{newtxtext,newtxmath}
\usepackage[T1]{fontenc}

\DeclareRobustCommand{\VAN}[3]{#2}
\let\VANthebibliography\thebibliography
\def\thebibliography{\DeclareRobustCommand{\VAN}[3]{##3}\VANthebibliography}

\usepackage{graphicx}	% Including figure files
\usepackage{amsmath}	% Advanced maths commands
\newcommand{\um}{~$\mu$m}
\newcommand{\kms}{km~s$^{-1}$}

\newcommand{\msun}{M$_{\odot}$}
\newcommand{\degree}{$^{\circ}$}
\newcommand{\tco}{$^{13}$CO}        %%"thirteen"CO, can't use number in command name
\title[CO in the Draco Nebula]{CO in the Draco Nebula:  The Atomic-Molecular Transition}

\author[J. H. Bieging and Shuo Kong]{
John H. Bieging$^{1}$\thanks{E-mail: jbieging@arizona.edu} and Shuo Kong$^{1}$
\\
$^{1}$Steward Observatory, The University of Arizona, Tucson AZ 85721  USA
}

\date{Accepted XXX. Received YYY; in original form ZZZ}

\pubyear{2024}

\begin{document}
\label{firstpage}
\pagerange{\pageref{firstpage}--\pageref{lastpage}}
\maketitle

\begin{abstract}

This paper presents maps of the J=2-1 transition of CO toward the Draco Nebula Intermediate Velocity Cloud (IVC).  The maps cover 8500 square arcmin with a velocity resolution of 0.33 \kms~ and angular resolution of 38\arcsec, or 0.11 pc at the cloud distance of 600 pc.  The mapped area includes all the emission detected by the {\it Herschel} satellite with 250 \um~ intensity >5 MJy/sr.  Previously published observations of the far-IR emission and the 21 cm line of HI are used to derive the column density distribution of H$_2$ and the abundance ratio CO/H$_2$, as well as the distribution of the molecular fraction of hydrogen, which approaches 90\% over much of the brighter parts of the nebula.  The CO emission is highly clumpy and closely resembles the structures seen in far-IR images.  The kinematics of the CO show supersonic motions between clumps but near-thermal to trans-sonic motions within clumps, consistent with model predictions that the scale length for dissipation of supersonic turbulence should be $\sim0.1$ pc, mediated by kinematic viscosity and/or ambipolar diffusion. 
 Different parts of the nebula show evidence for a spread of molecular formation timescales of a few 10$^5$ years, comparable to the dynamical timescale of the infalling gas.  The IVC will likely merge with the Galactic interstellar medium in $\sim 10^7$ years, and the densest clumps may form an unbound cluster of low-mass stars.

\end{abstract}


%%%%%%%%%%%%%%%%%%%%%%%%%%%%%%%%%%%%%%%%%%%%%%%%%%

%%%%%%%%%%%%%%%%% BODY OF PAPER %%%%%%%%%%%%%%%%%%

\section{Introduction}

This paper presents new observations combined with previously published results to address one of the key questions about the processes that occur in the interstellar medium, namely how neutral atomic gas is converted to the cold, dense molecular clouds which are the necessary preconditions for star formation.  It has long been recognized that the gas between the stars is not confined to a quiescent layer in the disk of the Milky Way.  Early observational studies of the 21 cm line of neutral atomic hydrogen in both emission and absorption showed that there are two dominant regimes of temperature which exist in approximate pressure equilibrium.  These ``phases" are generally referred to as the warm neutral medium (WNM) with a typical kinetic temperature of $\sim 8000$~K, and the cold neutral medium (CNM) with temperatures of $\sim 80$~K.  The physical explanation for these components was first given by \citet{1969ApJ...155L.149F}, who showed that the shape of the cooling function for the interstellar gas allowed the simultaneous existence of two stable temperature regimes in approximate pressure equilibrium.  Further studies of thermal bremsstrahlung emission at X-ray wavelengths, as well as far-UV spectral lines from highly ionized elements revealed a third component, the diffuse hot intercloud medium (HIM) with temperatures of $\sim 10^6$~K, that appears to be pervasive within the galactic disk.  The hot gas likely results from the collective effects of supernovae explosions and the winds of massive stars.  Since the cooling timescale for gas at this temperature is very long, the HIM can coexist with the other phases for at least millions of years, as shown by \citet{1977ApJ...218..148M}.

An additional component of ionized gas at a temperature of $\sim 8000$~K, the warm ionized medium or WIM, has also been identified by various observational diagnostics such as hydrogen recombination lines  and pulsar dispersion measures, but for the object considered in this paper, there appears to be very little WIM along the line of sight.  Therefore we will concentrate on the role of the WNM, CNM, and the formation of dense  molecular clouds from the atomic gas.  

The HIM may be part of the story at least indirectly through the action of supernovae in the distant past which created the conditions for a ``galactic fountain".  This concept was first proposed to explain the existence of the High Velocity Clouds (HVCs) detected in early HI 21 cm line surveys, which revealed discrete clouds of neutral atomic hydrogen with large, mostly negative radial velocities.  The number and inferred masses implied that these clouds were falling onto the Galactic disk with an improbably large mass flux \citep{1976ApJ...205..762S}.  An alternative explanation was that the infalling gas had previously been part of the Galactic disk gas and was blown into the Galactic halo by supernova explosions \citep{1980ApJ...236..577B}.  Eventually some of the gas cooled and condensed as a result of thermal instability and was drawn back toward the disk by the gravitational potential of the stars.  Further studies of the HI emission at high galactic latitudes showed that the clouds with negative radial velocities (i.e., moving in toward the Galactic disk) tended to fall into two categories:  those with velocities more negative than about $-100$~ \kms, referred to as high velocity clouds (HVCs); and those with radial velocities in the range of roughly $-100$~ \kms~ to $-20$~ \kms, referred to as intermediate velocity clouds (IVCs).  HI emission features with radial velocities near zero were attributed to low-velocity neutral atomic clouds (``LVC") within the Galactic disk and participating in differential galactic rotation.  

The object of the present study, the Draco Nebula (hereafter, ``Draco"), is considered an archetype of the IVC category.  In Galactic coordinates the center of the nebula is near $l = 91$\degree,~ $b = 38$\degree.  The overall extent of the nebula is about $4$\degree $\times 4$\degree.   The distance to the IVC has been estimated by various authors;  in this paper, we follow \citet{2017A&A...599A.109M} and adopt a distance of 600 pc, which places the nebula about 370 pc above the Galactic plane, with an overall size of $\sim 40$~pc.  The Draco IVC is therefore within the inner Galactic halo, somewhat above the bulk of the WNM in the disk \citep{1990ARA&A..28..215D}.  For a more extensive description of the properties as well as a chronology of observational studies of Draco, see the paper by \citet{2017A&A...599A.109M} which also motivated our study of CO emission.  

Early surveys of CO J=1-0 emission \citep{1985A&A...151..427M,1989A&A...211..402R} showed that Draco had a significant amount of molecular gas for an IVC HI cloud, compared with other IVCs.  This anomaly suggested that some fraction of the neutral atomic gas had been cooled and condensed sufficiently to promote the formation of molecules,  but the details of this process were unclear.  In  observations with the {\it Herschel} far-IR space telescope, however, \citet{2017A&A...599A.109M} found that the thermal dust continuum (which they mapped in the 250 $\mu$m band with the SPIRE instrument), showed numerous narrow structures extending toward the Galactic plane.  These "fingers" of emission exhibited features down to the resolution limit of the image (18\arcsec) and some were probably unresolved.  \citet{2017A&A...599A.109M} interpreted these structures as condensations of the HI gas in the IVC resulting from a Rayleigh-Taylor type instability as the cold, denser HI cloud encountered the WNM of the Galactic disk while falling back into the Galaxy as part of the "galactic fountain" scenario.  
If this picture were correct, then the 250 $\mu$m structures ought to coincide with CO emission, assuming that the cold, dense features marked the regions where the neutral atomic gas was turning into a molecular state.  The {\it Herschel} observations, however, showed only the thermal dust continuum, not the presence of molecular gas.   An earlier study by \citet{1993A&A...272..514H} revealed concentrations of CO J=1-0 emission that did coincide with the brighter peaks of the {\it Herschel} 250 $\mu$m image, but the sensitivity and areal coverage of the CO map were limited to the brightest parts of the far-IR emission.  We therefore undertook a more extensive mapping program, described in the next section, to look for quantitative relations between the thermal dust continuum (as a tracer of total hydrogen column density), the atomic hydrogen content as traced by the 21 cm emission line, and the molecular gas traced by CO emission.  

\section{Observations}

\subsection{CO J=2-1 mapping}
\label{sec:CO21} % used for referring to this section from elsewhere

All the new observations of CO J=2-1 emission presented in this paper were obtained with the Heinrich Hertz Submillimeter Telescope\footnote{For technical specifications, see the website https://aro.as.arizona.edu/?q=facilities/submillimeter-telescope}~on Mt. Graham, Arizona (elevation 3200 m).  The observations used the on-the-fly (OTF) method to obtain spatially fully-sampled maps of 10\arcmin~ square ``tiles", similar to other molecular cloud maps we have published previously.  See, e.g., \citet{Bieging_2022} and references therein for details of the mapping and calibration.  The final maps have been lightly smoothed to a resolution of 38\arcsec~(FWHM), or 0.11 pc at the adopted distance of 600 pc.  The spectral resolution was 250 kHz, which corresponds to 0.33 \kms~ at the line frequency of 230 GHz.  At this resolution, the typical rms noise level in one velocity channel is 0.1 K brightness temperature.  The total extent of the analyzed spectral coverage was from -43 to -8 \kms (Local Standard of Rest) at this resolution.  We also obtained spectral line data at 1 MHz resolution that covered $\sim 330$ \kms, which allowed us to search for CO emission in the velocity range of both the LVC and the HVC located in the general direction of Draco, as shown in the HI maps of \citet{2017A&A...599A.109M}.  No CO was detected in either the LVC or the HVC velocity range toward the main Draco features.

Our observations were motivated by the {\it Herschel} map presented by \citet{2017A&A...599A.109M}, which covers an area of about 4 degrees on a side.  A spatially complete map of CO over such a large area would have been prohibitively time-consuming.  From initial observations of some of the brightest far-IR features, however, it was clear that CO is only detectable where the 250 $\mu$m surface brightness exceeds about 5 MJy/steradian.  We therefore placed the mapping tiles to cover only regions that met this criterion.  The CO maps follow the far-IR structures in an irregular pattern.  In total, we mapped some 85 tiles or about 8500 square arcminutes.  Figure \ref{fig:Fig1_4combo}(a) shows the {\it Herschel} map from \citet{2017yCat..35990109M}, convolved to a resolution of 38\arcsec~ to match the final resolution of our CO maps.  The labelled white rectangles identify subregions that will be considered in detail below;  we use these designations for convenience to refer to specific areas of the nebula.  
Figure \ref{fig:Fig1_4combo}(b) shows the integrated CO J=2-1 brightness temperature in Kelvins-\kms, for all of the mapped tiles.  The zero level is offset into the dark blue of the color palette, so where CO is not detected, the dark blue background illustrates both the extent of the mapped tiles, and the generally uniform noise levels achieved with these observations.  The white contour shows the 5 MJy/sterad intensity level of the 250 $\mu$m emission from Figure \ref{fig:Fig1_4combo}(a).  It is evident that the CO emission is entirely confined to the sight-lines where the 250 $\mu$m surface brightness exceeds 5 MJy/sterad.  The CO emission at the top of the map extends slightly beyond the white contour, because the {\it Herschel} map is cut off at that position---see Figure~\ref{fig:Fig1_4combo}(a)---even though there is detectable CO beyond the far-IR map edge.

As with our previous studies of Galactic Giant Molecular Clouds (GMCs), we also mapped the \tco~ J=2-1 transition simultaneously.  In contrast to typical GMCs, however, the \tco~ line in Draco is mostly undetected except at the very brightest CO emission peaks.  We will therefore not include the mapping data for the \tco~ line in this paper.  Where \tco~ was detected, the ratio of line intensities (CO/\tco) will be used to estimate the optical depth of the CO J=2-1 line.

\subsection{Published observations used in the analysis}
\label{Published observations}

Critical to this study are previously published observations which we use to derive:  the distribution of total column density of hydrogen in all forms; the distribution of neutral atomic hydrogen in the radial velocity range of the IVC; and from those, to infer the column density distribution of molecular hydrogen in the IVC.  

\begin{figure*}
    \includegraphics[width=3.2in]{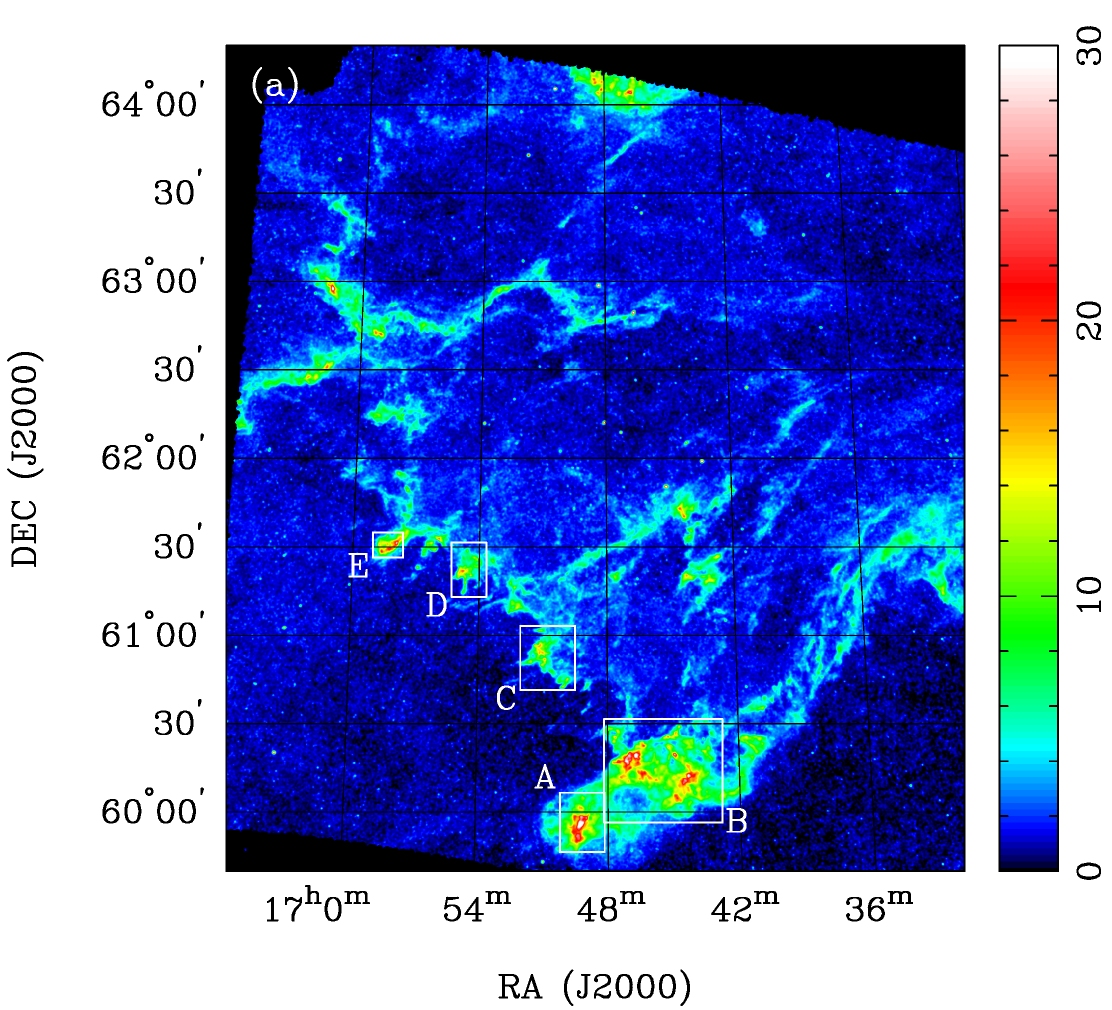}
    \includegraphics[width=3.2in]{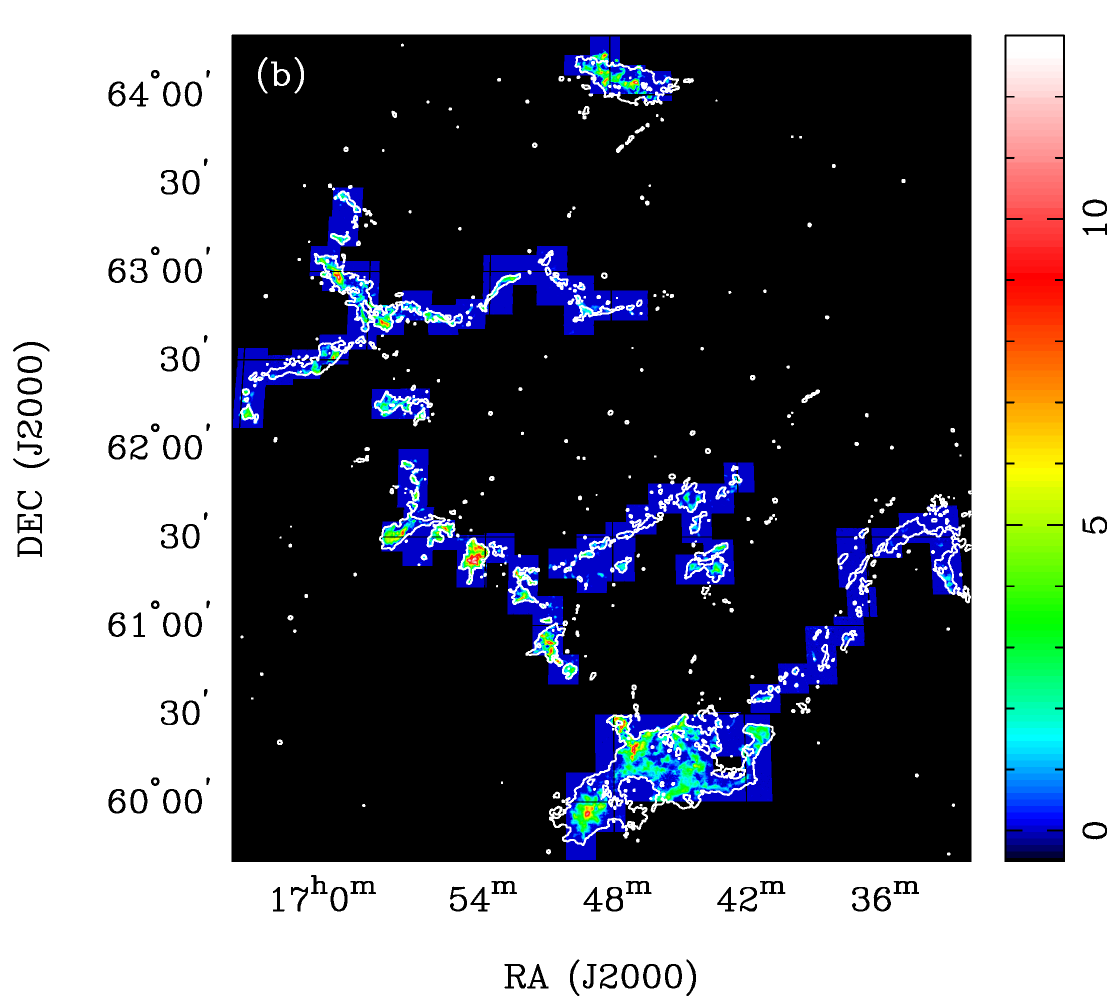}\\
    \includegraphics[width=3.2in]{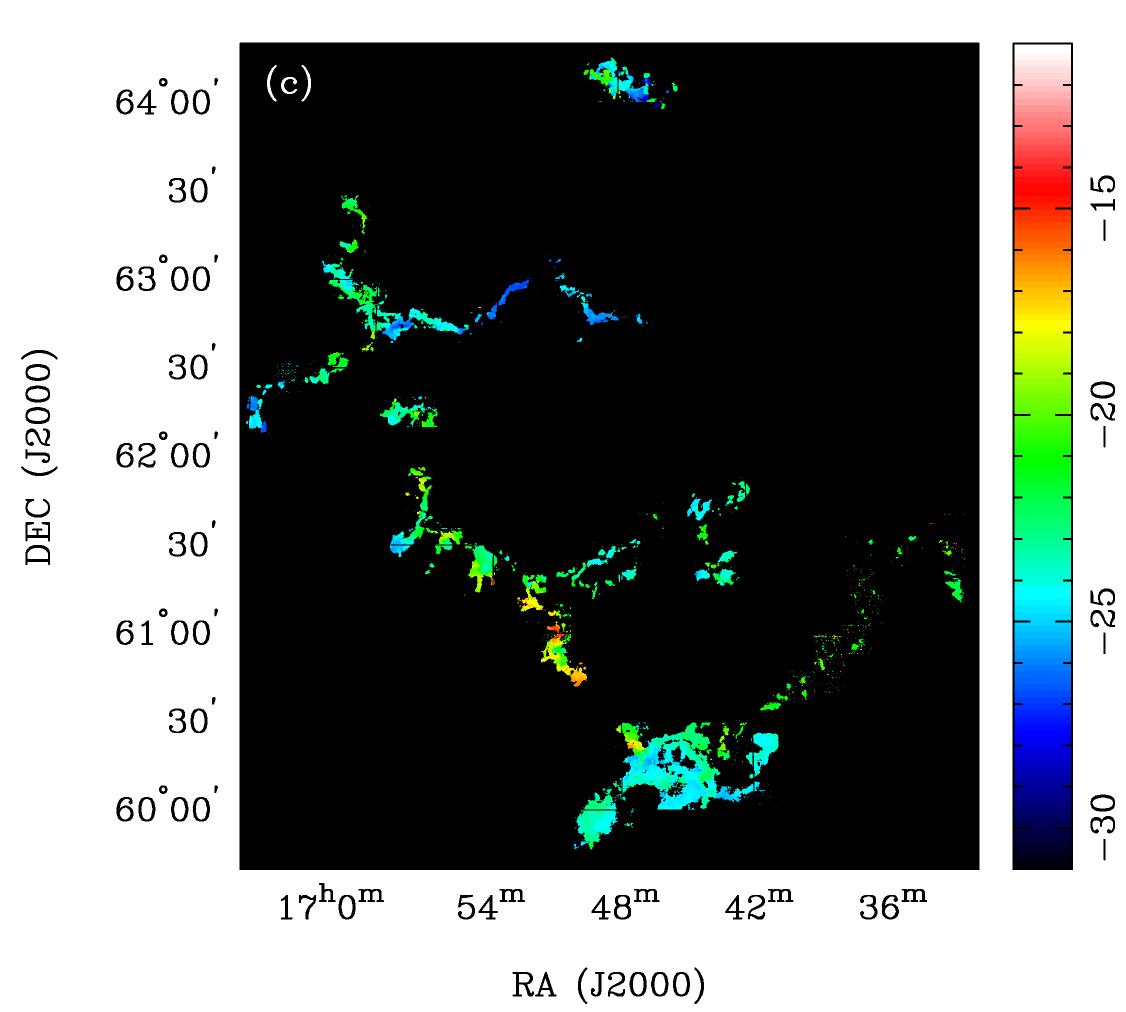}
    \includegraphics[width=3.2in]{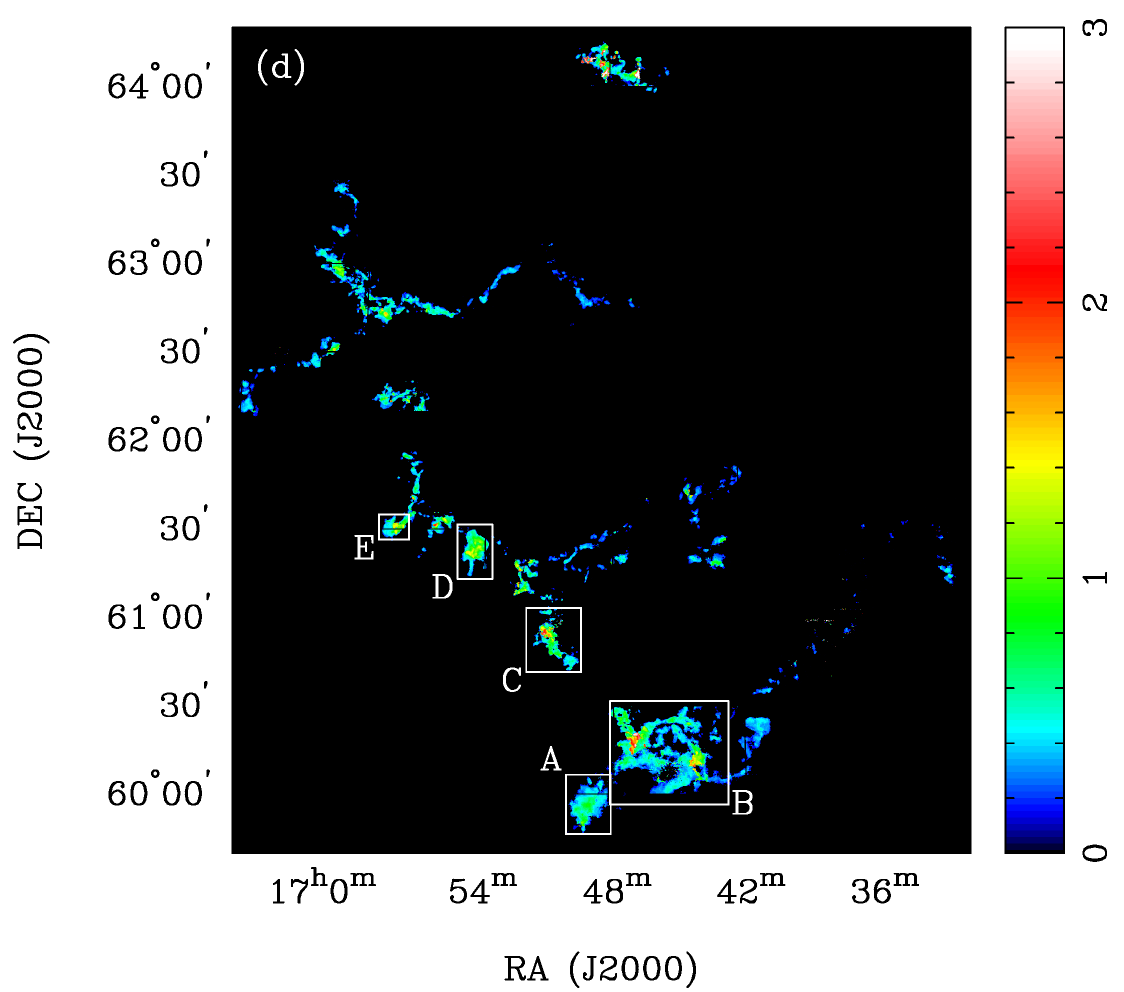}
    \caption{(a) {\it Herschel} 250 $\mu$m map {\citep{2017yCat..35990109M}} smoothed to a resolution of 38\arcsec~ (FWHM) to match the CO maps.  Color wedge is labelled in MJy/sterad, with a square-root stretch to enhance the lower surface brightness features.  Labelled boxes show selected areas discussed in Section \ref{sec:COmaps_and_moments}.  (b) Whole field CO J=2-1 integrated brightness temperature over the IVC velocity range.  White contour shows {\it Herschel} 250 $\mu$m continuum at 5 MJy/sterad.  Resolution is 38\arcsec~ (FWHM).  Color wedge is labelled in K-\kms, with a square-root stretch to enhance the lower surface brightness features.  No pixels in the CO mapping fields were blanked, so the dark blue backgrounds show the extent of the OTF-mapped tiles.  (c)  Whole field velocity centroid (first moment) in \kms~ (LSR).  (d)  Whole field velocity dispersion, $\sigma_V$ (second moment) in \kms.  For a Gaussian line shape, FWHM = $2.355 \sigma_V$. The labelled white rectangles show the areas selected for display of velocity channel maps (Figures \ref{fig:BoxAchanmaps} -- \ref{fig:BoxEchanmaps}) which illustrate the complexity of the molecular gas distribution in radial velocity.}
    \label{fig:Fig1_4combo}
\end{figure*}

To derive the distribution of total hydrogen in all forms, we use the {\it Herschel} 250~$\mu$m map downloaded from  \citet{2017yCat..35990109M} and apply the simple proportionality between far-IR surface brightness and gas column density prescribed in \citet{2017A&A...599A.109M}.  Their observations could not be used to derive the dust temperature distribution and from that, assuming a standard dust opacity law, the distribution of total gas column density.  They argue, however, that the lower resolution data from the {\it Planck} all-sky maps show that the thermal dust emission over this region has a very uniform temperature.  Then, assuming a standard dust/gas opacity law and the {\it Planck} dust temperature, they derive a linear proportionality which we use to convert their 250 $\mu$m map to a map of total hydrogen column density, at the resolution of the {\it Herschel} data, i.e., 18~\arcsec~(FWHM).  The {\it Herschel} I(250) surface brightness map (in MJy/sr), convolved to match the resolution of our CO data (38\arcsec) is shown in Figure \ref{fig:Fig1_4combo}(a).

We also considered two other, more recent analyses of the {\it Herschel} images to derive the dust temperature and opacity distributions \citep{Singh_2022,refId0}.  We concluded, however, that the uncertainties in critical parameters, namely the dust/gas mass ratio and the dust absorption coefficient, were so great that we opted to use the simple relation from \citet{2017A&A...599A.109M} to derive the total hydrogen column density.  The implications of the alternative dust models are discussed in Appendix \ref{Appendix}.

The distribution of the neutral atomic hydrogen can be derived more directly than the dust column density from observations of the HI 21 cm emission line.  \citet{2017A&A...599A.109M} also show maps of the HI 21 cm emission over three velocity ranges, comprising the HVC, IVC, and LVC, from the Green Bank survey of HI emission at intermediate galactic latitudes \citep{2015ApJ...809..153M}.  The angular resolution of this survey is $\sim 9.6$~\arcmin, i.e., much lower than the {\it Herschel} map, but the separation in velocity shows that there is very little HI in the velocity range of the LVC toward the Draco emission at 250~$\mu$m.  There are some moderately bright HI peaks in the HVC velocity range toward the center of Draco, but the dominant IVC HI emission is clearly well-correlated with the {\it Herschel} 250 $\mu$m map.  We examined our CO J=2-1 spectra at 1 MHz resolution over the velocity range of the HVC from some of our map tiles located near the HVC peak from \citet{2017A&A...599A.109M}--see their Figure 2, top panel.  There was no CO detected at those positions in the velocity range of the HVC, to a brightness temperature limit of $\sim 0.1$~K.  (The non-detection of CO in the HVC velocity range may indicate that the HVC is extragalactic in origin and of low metallicity.)  

A map of the HI 21 cm line over most of the Draco region was obtained with the DRAO hydrogen line interferometer array, which combined with the Green Bank survey, was used to produce the DHIGLS survey \citep{Blagrave_2017}, at an angular resolution of $\sim 1$\arcmin.  This survey provides HI maps separated into the LVC and IVC velocity ranges.  We have convolved the {\it Herschel} 250 $\mu$m image to match that of the DHIGLS survey (which covers most but not all of the {\it Herschel} field).  Then, assuming that the Draco nebula is dominated by the IVC gas, we subtract the DHIGLS HI column density map from the corresponding map of total hydrogen gas density to obtain a residual map which represents the distribution of molecular hydrogen, limited to the DHIGLS field and angular resolution. 

\section{Results and analysis}
\label{sec:results}
\subsection{CO J=2-1 maps and velocity moments}
\label{sec:COmaps_and_moments}

A map of the intensity of the CO J=2-1 line integrated over the velocity range of the IVC is shown in Figure \ref{fig:Fig1_4combo}(b) for the entire mapped region.   
As noted in Section 2.1, the \tco~ line is mainly undetected over the entire field which implies that the CO J=2-1 toward Draco is optically thin over the whole nebula.  Therefore the integrated intensity of CO should be a good indicator of the CO column density distribution.  In section \ref{sec:LTEanalysis}, we use an analysis that assumes LTE excitation conditions to derive quantitative values for the CO total column density and compare those with the column densities and spatial distributions of the molecular and neutral atomic hydrogen.  

In their study of Draco, \citet{2017A&A...599A.109M} show that the structures in the far-IR thermal dust emission are likely the result of a Rayleigh-Taylor (R-T) instability that develops as the infalling IVC atomic hydrogen cloud encounters the layer of WNM in the Galactic disk.  Such a dynamical process would be expected to show velocity effects as the infalling gas is decelerated upon encountering the WNM, likely with variations over the $\sim 40$~pc extent of the IVC.  The {\it Herschel} map is limited to the thermal dust continuum, however, so no velocity information can be derived from those data.  Our CO spectral line data cube has a resolution on the velocity axis of 0.33 \kms, allowing us to look in some detail at the velocity structure of the R-T condensations.  These appear to be the regions where the atomic gas is transforming to a predominantly molecular state, of which CO is a tracer.  
Figure \ref{fig:Fig1_4combo}(c) shows the velocity centroid (i.e., the intensity-weighted mean velocity) over the entire mapped field.  The velocities, referred to the Local Standard of Rest, range over some 12 \kms, from $-15$~\kms~ to $-27$~\kms.  The overall appearance of the velocity centroid map is not entirely chaotic;  along the sinuous east-west filament between declinations +62\degree 30\arcmin~ and +63\degree, there appear to be two sections with roughly constant centroid velocities, but with a sharp boundary between them across an apparent break in the CO emission, amounting to a difference of 3 \kms~($-27$ vs. $-24$ \kms).  The most typical velocity centroid value is -24 \kms.  The region extending along the southeast side of the nebula, which is where \citet{2017A&A...599A.109M} find the best examples of R-T condensations, has generally less negative centroid velocities, possibly consistent with a deceleration of the infalling gas as it encounters the Galactic layer of WNM.  

\begin{figure*}
	\includegraphics[width=5.5in]{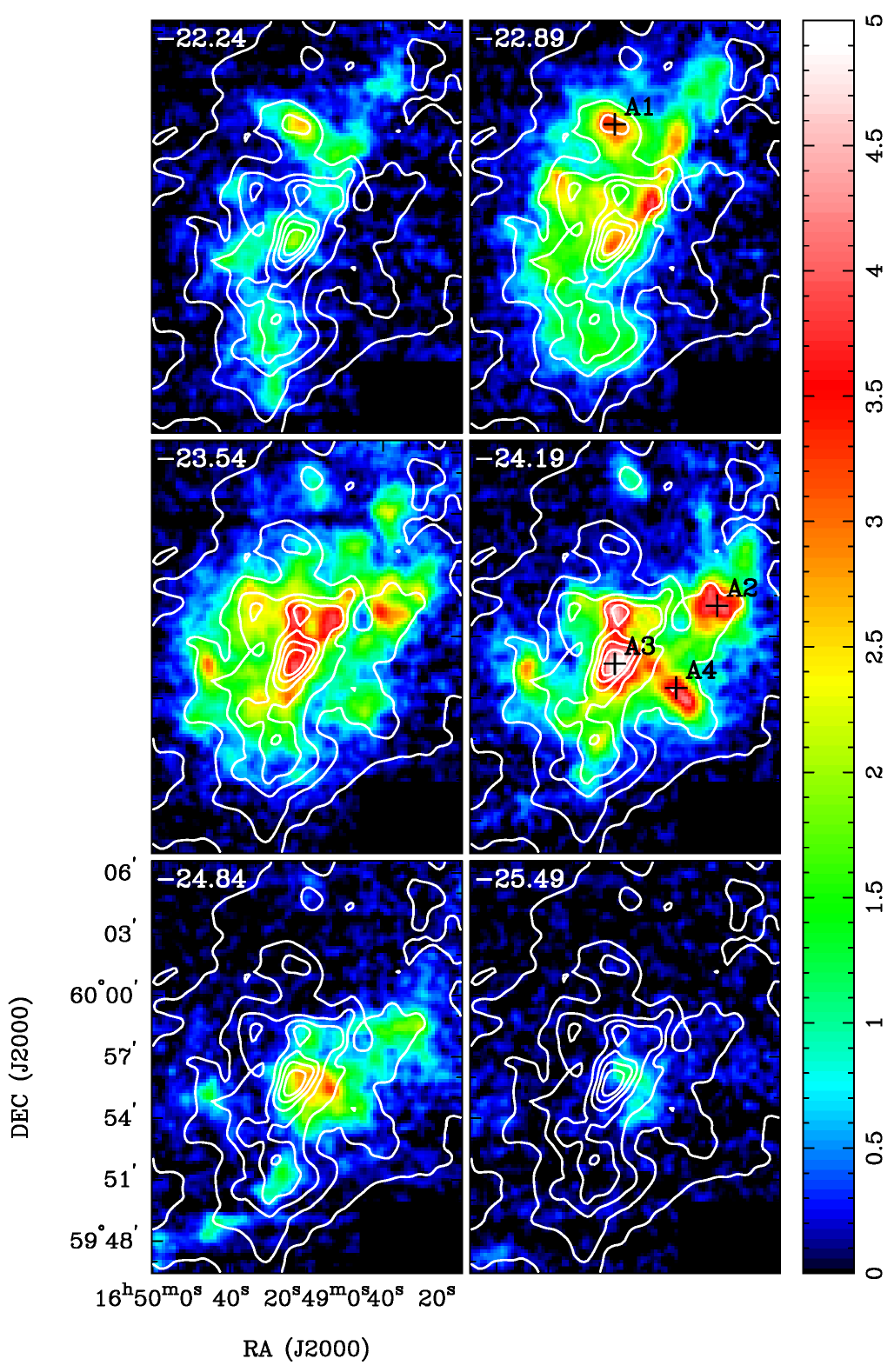}
    \caption{Channel maps for selected region A in Figure \ref{fig:Fig1_4combo}.  LSR velocity is in upper left of each panel;  every 2nd channel is shown.  Color wedge is in Kelvins brightness temperature. White contours show the 250 $\mu$m surface brightness from the {\it Herschel} map convolved to 38\arcsec~ resolution (see Figure \ref{fig:Fig1_4combo}a);  contour levels are 5,  10, 15, 20, 25, 30, and 35 MJy/sterad.  Labelled crosses mark positions of spectra shown in Fig. \ref{fig:SpecPlotsSelectedRegions}, and are shown only in the velocity channel map nearest the peak of the spectrum.  Labels are numbered in order of decreasing declination.}
    \label{fig:BoxAchanmaps}
\end{figure*}

\begin{figure*}
	\includegraphics[width=6.5in]{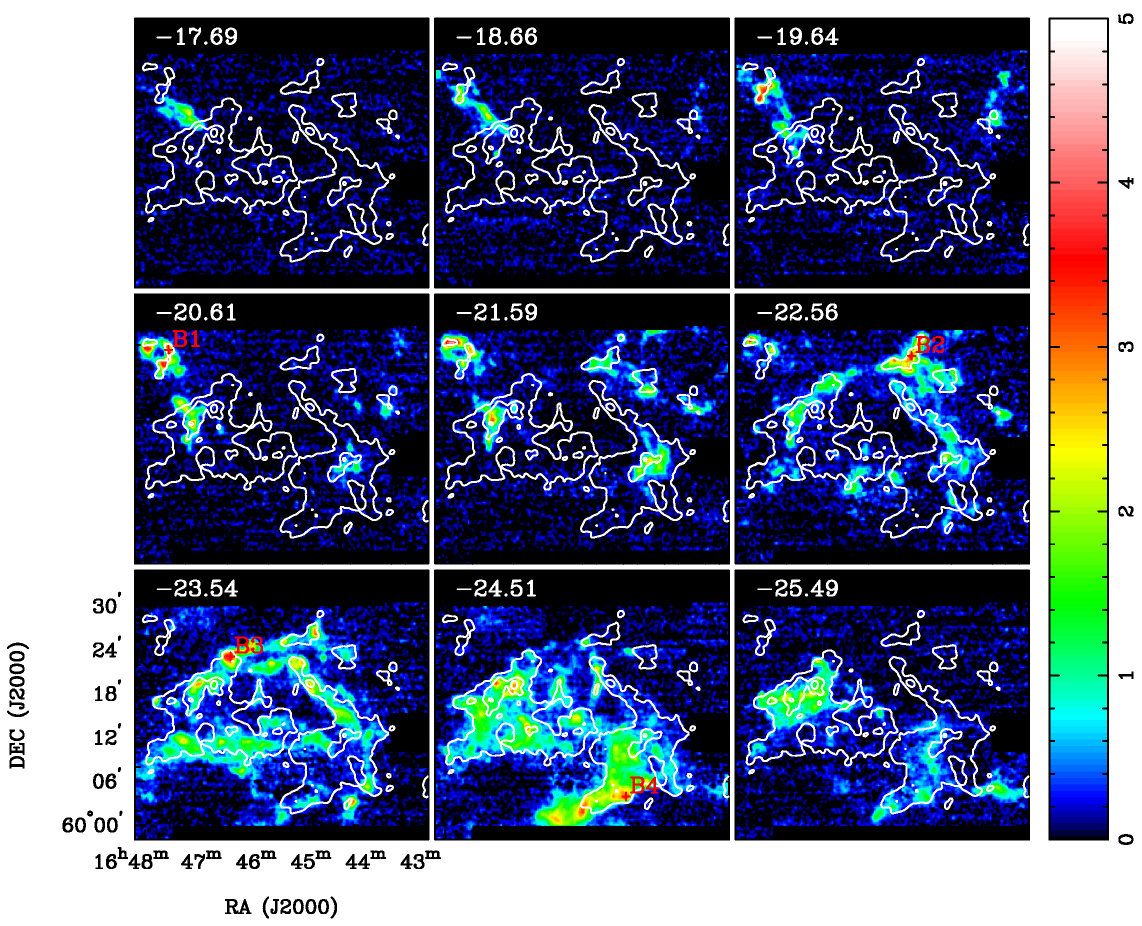}
    \caption{Channel maps for selected region B in Figure \ref{fig:Fig1_4combo}.  LSR velocity is in upper left of each panel;  every 3rd channel is shown.  Color wedge is in Kelvins brightness temperature. White contours show the 250 $\mu$m surface brightness from the {\it Herschel} map convolved to 38\arcsec~ resolution (see Figure \ref{fig:Fig1_4combo}a);  contour levels are 10, 20 and 30 MJy/sterad.  Labelled crosses mark positions of spectra shown in Fig. \ref{fig:SpecPlotsSelectedRegions}, and are shown only in the velocity channel map nearest the peak of the spectrum.  Labels are numbered in order of decreasing declination.}
    \label{fig:BoxBchanmaps}
\end{figure*}

\begin{figure*}
	\includegraphics[width=6.5in]{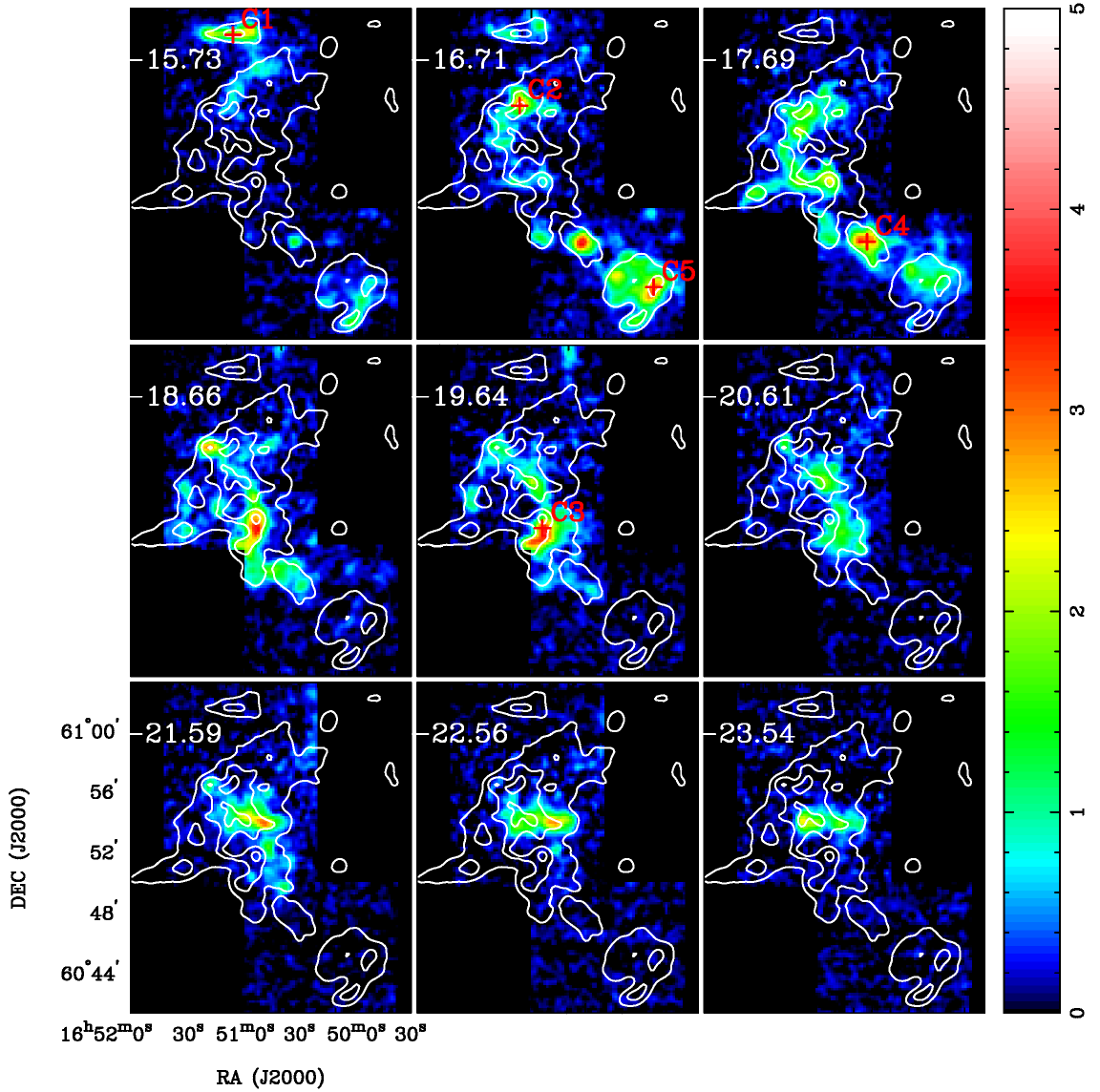}
    \caption{Channel maps for selected region C in Figure \ref{fig:Fig1_4combo}.  LSR velocity is in upper left of each panel;  every 3rd channel is shown.  Color wedge is in Kelvins brightness temperature.  White contours show the 250 $\mu$m surface brightness from the {\it Herschel} map convolved to 38\arcsec~ resolution (see Figure \ref{fig:Fig1_4combo}a);  contour levels are 5, 10, and 15 MJy/sterad.  Labelled crosses mark positions of spectra shown in Fig. \ref{fig:SpecPlotsSelectedRegions}, and are shown only in the velocity channel map nearest the peak of the spectrum.  Labels are numbered in order of decreasing declination.}
    \label{fig:BoxCchanmaps}
\end{figure*}

\begin{figure*}
	\includegraphics[width=6.0in]{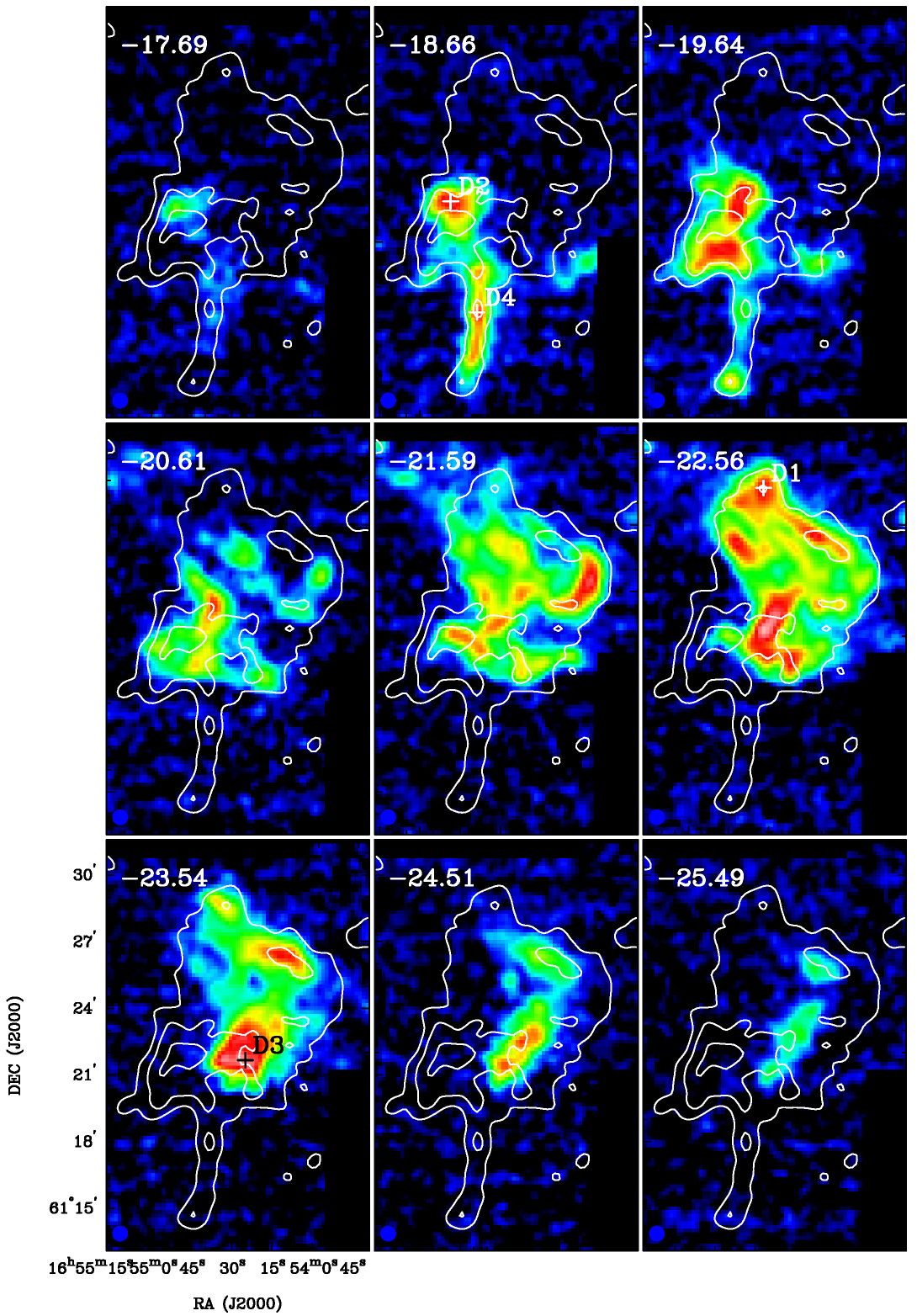}
    \caption{Channel maps for selected region D in Figure \ref{fig:Fig1_4combo}.  LSR velocity is in upper left of each panel;  every 3rd channel is shown.  Color wedge is in Kelvins brightness temperature.  White contours show the 250 $\mu$m surface brightness from the {\it Herschel} map convolved to 38\arcsec~ resolution (see Figure \ref{fig:Fig1_4combo}a);  contour levels are 5, 10, and 15 MJy/sterad.  Labelled crosses mark positions of spectra shown in Fig. \ref{fig:SpecPlotsSelectedRegions}, and are shown only in the velocity channel map nearest the peak of the spectrum.  Labels are numbered in order of decreasing declination.}
    \label{fig:BoxDchanmaps}
\end{figure*}

\begin{figure*}
	\includegraphics[width=6.5in]{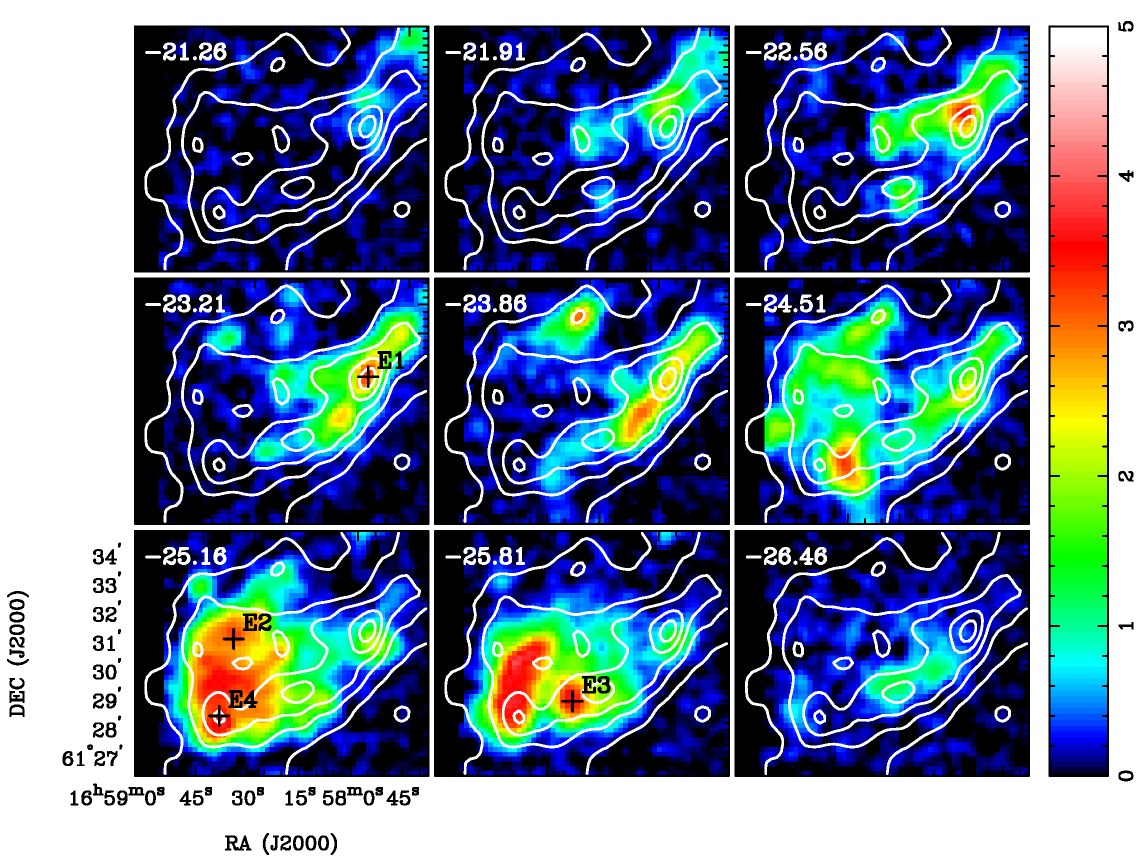}
    \caption{Channel maps for selected region E in Figure \ref{fig:Fig1_4combo}.  LSR velocity is in upper left of each panel;  every 2nd channel is shown.  Color wedge is in Kelvins brightness temperature.  White contours show the 250 $\mu$m surface brightness from the {\it Herschel} map convolved to 38\arcsec~ resolution (see Figure \ref{fig:Fig1_4combo}a);  contour levels are 5, 10, 15, 20, and 25 MJy/sterad.  Labelled crosses mark positions of spectra shown in Fig. \ref{fig:SpecPlotsSelectedRegions}, and are shown only in the velocity channel map nearest the peak of the spectrum.  Labels are numbered in order of decreasing declination.}
    \label{fig:BoxEchanmaps}
\end{figure*}

\begin{figure*}
    \includegraphics[width=6.5in]{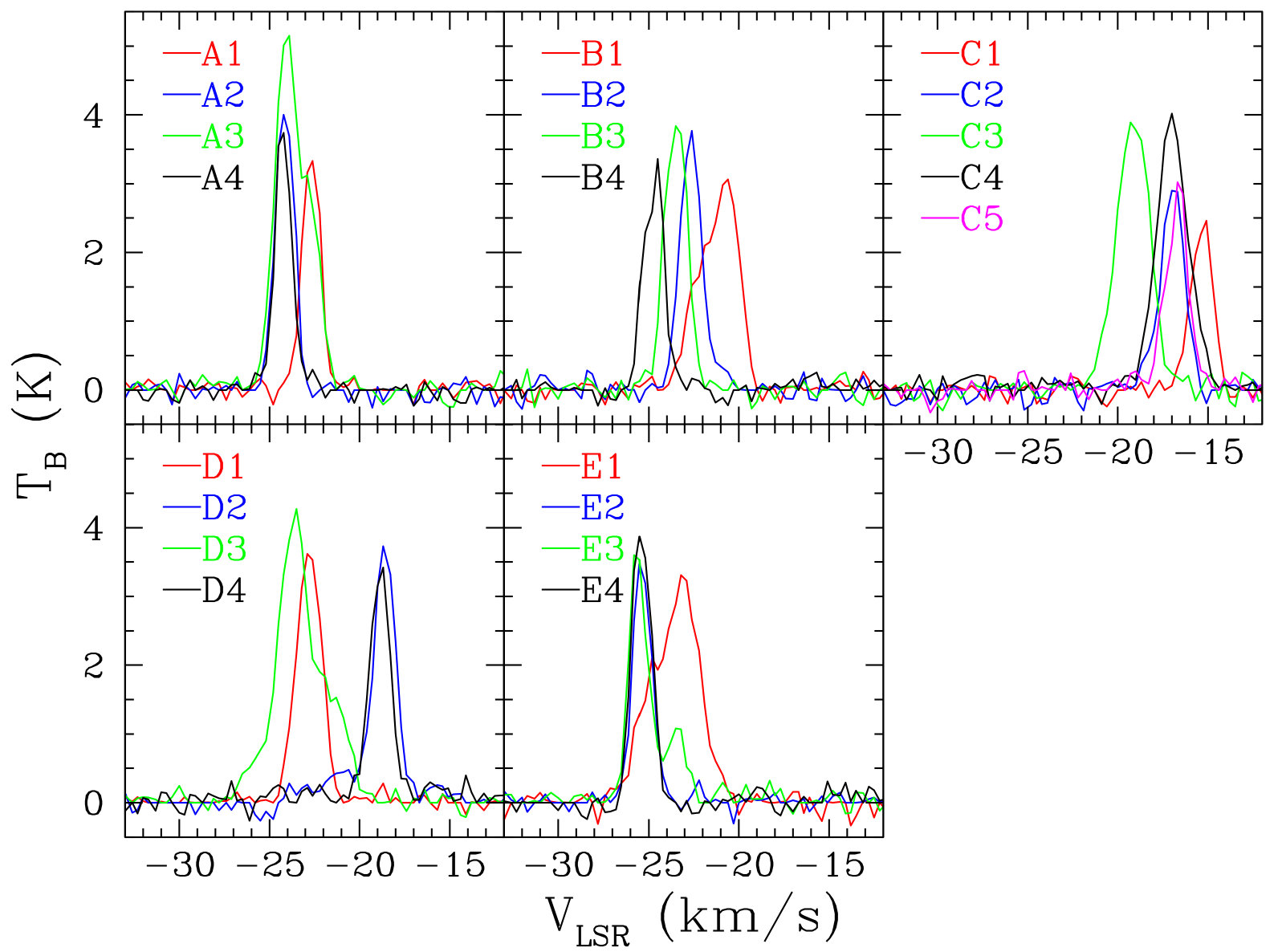}
    \caption{ Spectra for single pixels in selected regions A - E, shown in Figures \ref{fig:BoxAchanmaps} - \ref{fig:BoxEchanmaps}.  Spectra are numbered in order of decreasing declination.  The effective angular resolution is 38\arcsec~ (FWHM) and the velocity resolution is 0.33 \kms.} 
    \label{fig:SpecPlotsSelectedRegions}
\end{figure*}

In Figure \ref{fig:Fig1_4combo}(d) we show the distribution of the second velocity moment or dispersion, $\sigma_V$, of the CO line.  (If the line shape is a Gaussian, then the full width at half maximum intensity equals $2.355 \sigma_V$.)  The most common velocity dispersion is $\sigma_V \approx 0.35$~\kms~as shown in a histogram of the distribution of $\sigma_V$.  The sound speed for the gas at T=20 K is about 0.3 \kms, so the internal motions of the molecular gas are generally not much above thermal.  There is a tail of values in the distribution extending up to $\sim 3$~\kms.  Some of the largest line widths appear near the brightest CO emission peaks, particularly in the southeast and south.  These areas are likely where the R-T instability is most dominant according to \citet{2017A&A...599A.109M}.  The instability may lead to increased line broadening via turbulence and multiple density components along the line of sight.  The narrowest lines are generally in regions located away from the R-T features identified by \citet{2017A&A...599A.109M}, i.e., the dark blue pixels in Figure \ref{fig:Fig1_4combo}(d).  In the following paragraphs, we examine spectral channel maps (Figures \ref{fig:BoxAchanmaps} - \ref{fig:BoxEchanmaps}) and line profiles (Figure \ref{fig:SpecPlotsSelectedRegions}) in some of these R-T features. 

To illustrate the complexity of the velocity field in the CO emission, we selected 5 regions labelled A - E, as shown in Figure \ref{fig:Fig1_4combo}(a) and (d), to display individual channel maps of the CO emission with spectral resolution of 0.33 \kms~ (which is about the same as the sound speed in the gas at T=20 K.)  These maps are in Figures \ref{fig:BoxAchanmaps} -- \ref{fig:BoxEchanmaps};  every 2nd or 3rd channel is shown, as indicated in the figure captions.   Regions C, D, and E, and the upper left corner of region B, correspond to the R-T instability structures shown in Figure 4 of \citet{2017A&A...599A.109M}.  The CO channel maps have a spatial resolution of 38\arcsec~ while the original {\it Herschel} map in \citet{2017A&A...599A.109M} has 18\arcsec~ resolution.  In the channel map figures we show as white contours the {\it Herschel} 250 $\mu$m map smoothed to match the CO map resolution.  Also shown are labelled crosses that mark the pixels where CO spectra are displayed in Figure \ref{fig:SpecPlotsSelectedRegions}. (The crosses are only marked in the map of the channel nearest the emission peak of each spectrum.)  The CO channel maps clearly show that there is a complex velocity structure within the R-T instability "fingers'', even when the far-IR map is smoothed to 38\arcsec~ resolution.  Here we comment briefly on the velocity structure in each of the five selected fields:

    --Field A (Fig. \ref{fig:BoxAchanmaps}):  This field is the southern-most feature in Draco.  The {\it Herschel} 250 $\mu$m map has several clumpy structures as shown in the white contours, and the CO channel maps make clear that these clumps have a spread of velocities, with some of the far-IR peaks showing CO emission in only one or two of the 0.33 \kms channels.   Thus, the CO line width in each of the clumps is narrower than the spread of velocities over the entire field.  

    --Field B (Fig. \ref{fig:BoxBchanmaps}):  This field contains a complex set of far-IR structures, and together with Field A, dominates the {\it Herschel} map of Draco (e.g., Fig. \ref{fig:Fig1_4combo}(a)).  The top left corner of the field is included in Figure 4 of \citet{2017A&A...599A.109M}, which illustrates the features they identify as resulting from the R-T instability.  The CO channel maps show that this gas has radial velocities between about $-18$ and $-22$ \kms ~(LSR), while the more prominent far-IR emission peaks in the east-central and west-central parts of the field have velocities between about $-22$ and $-26$ \kms.  As in Field A, the many clumps generally appear in only a few CO channels, implying that the individual components have relatively narrow CO line widths, on an overall larger spread in systemic velocity.
    
    --Field C (Fig. \ref{fig:BoxCchanmaps}):  In the high-resolution Figure 4 in \citet{2017A&A...599A.109M}, this feature shows numerous compact clumps of dust emission.  As with Fields A and B,  the CO maps for Field C show that the dust clumps have radial velocities from about $-15$ to $-24$ \kms, but the individual components appear in only a few CO velocity channels.  There is no obvious large-scale organization to the clump velocities, suggesting that this field has a fairly chaotic velocity field rather than a more organized overall dynamical structure.  %A chaotic velocity field might be expected if the R-T "fingers`` are dominated by turbulence.
    
    --Field D (Fig. \ref{fig:BoxDchanmaps}):  This feature is one of those resulting from the R-T instability according to \citet{2017A&A...599A.109M}.  Their {\it Herschel} map shows a clumpy knot with a narrow extension to the south.  Our CO channel maps show that this field, like the others, has many bright CO peaks which appear in only a few velocity channels, so each peak has a CO line width of order 1 \kms, while the overall velocity range of the emission extends from $-17$ to $-26$ \kms.  The narrow "tail`` extending to the south of the main emission appears mainly in only one channel, so the molecular gas in the narrow "tail`` must have an internal velocity spread  <1 \kms.  
    
    --Field E (Fig. \ref{fig:BoxEchanmaps}):  This feature is prominent in the {\it Herschel} map, with a clumpy, almost cometary appearance.  The far-IR emission shows a bright area stretched along the southwest side and bright head consisting of a clumpy structure at the southeast end.  The CO channel maps show a more organized velocity structure compared to the other fields, in that the upper right (northwest) corner has an LSR velocity around $-23$ \kms, with a gradient to the southeast.  The brightest part of the far-IR image has a radial velocity in the range $-25$ to $-26$ \kms.  At the intermediate velocity around $-24$ \kms, there appear to be two extended features stretching from the southeast to the northwest, suggestive of gas being ablated and decelerated from the main clump as it plunges into the WNM layer.  Such a scenario is admittedly somewhat speculative but is suggested by the interpretation of a R-T instability forming the observed structures.  It would be of interest to see if numerical hydrodynamic simulations could reproduce both the density and velocity structures that are observed.
    
Spectra at the individual pixels marked in Figures \ref{fig:BoxAchanmaps} - \ref{fig:BoxEchanmaps} are displayed in Figure \ref{fig:SpecPlotsSelectedRegions}.  Many of the spectra have FWHM line widths of $\sim 1$ \kms (e.g., A1, A2, A4, E2, E4), while the broader lines appear to consist of two or more individual components seen superimposed along the line of sight (e.g., A3, E1, E3).  A detailed analysis of the entire Draco CO data cube for turbulence-related properties is beyond the scope of this paper but will be a subject for a future study.

\subsection{LTE analysis}
\label{sec:LTEanalysis}
The physical conditions of the molecular gas traced by CO emission are not well-constrained, so we must make assumptions to derive quantities such as total CO column density and abundance relative to molecular hydrogen.  In particular, a non-LTE approach using a Large Velocity Gradient treatment of radiative transfer would have required many poorly constrained assumptions.  Instead we opted for a simple LTE analysis, following the method described in \citet{Kong_2015}, to estimate the total column density of CO along each line of sight.  We assumed that the gas temperature was 20 K, i.e., the same as that used by \citet{2017A&A...599A.109M} for the nebular dust temperature.  We caution, however, that in the less dense parts of the nebula, the gas and dust may not be in thermal equilibrium so the gas kinetic temperature is not 20 K.  Fortunately, if the CO rotational level populations are close to LTE conditions, the total column density we derive from the J=2-1 transition is relatively insensitive to the assumed temperature.  For the observed transition, the line optical depth and the source function have roughly opposite temperature dependencies over the range of plausible values, i.e., between 10 K and 60 K the derived total CO column density varies by only $\pm 20\%$.  Thus, we believe that T$_{gas} = 20$~K is a reasonable guess for the LTE analysis.  See for example Appendix A in \citet{10.1111/j.1365-2966.2011.19279.x} for a comparison of the temperature dependence of the emissivity for the 3 lowest rotational lines of CO.

Assuming LTE applies and using the method in \citet{Kong_2015}, we find that the maximum optical depth of the CO J=2-1 line is 0.65, and the average over all detected pixels is 0.14 assuming a kinetic temperature of 20 K.  Therefore, the line is well described as optically thin, which should allow a robust calculation of total CO column density.  This low optical depth is corroborated  by the lack of \tco~ J=2-1 emission over virtually all of Draco.  If we assumed a lower kinetic temperature, e.g., as low as 10 K, the optical depths would increase by only about a factor of 2, hence still optically thin.  An assumed temperature higher than 20 K would result in even lower optical depths. 

\begin{figure*}
    \includegraphics[width=3.2in]{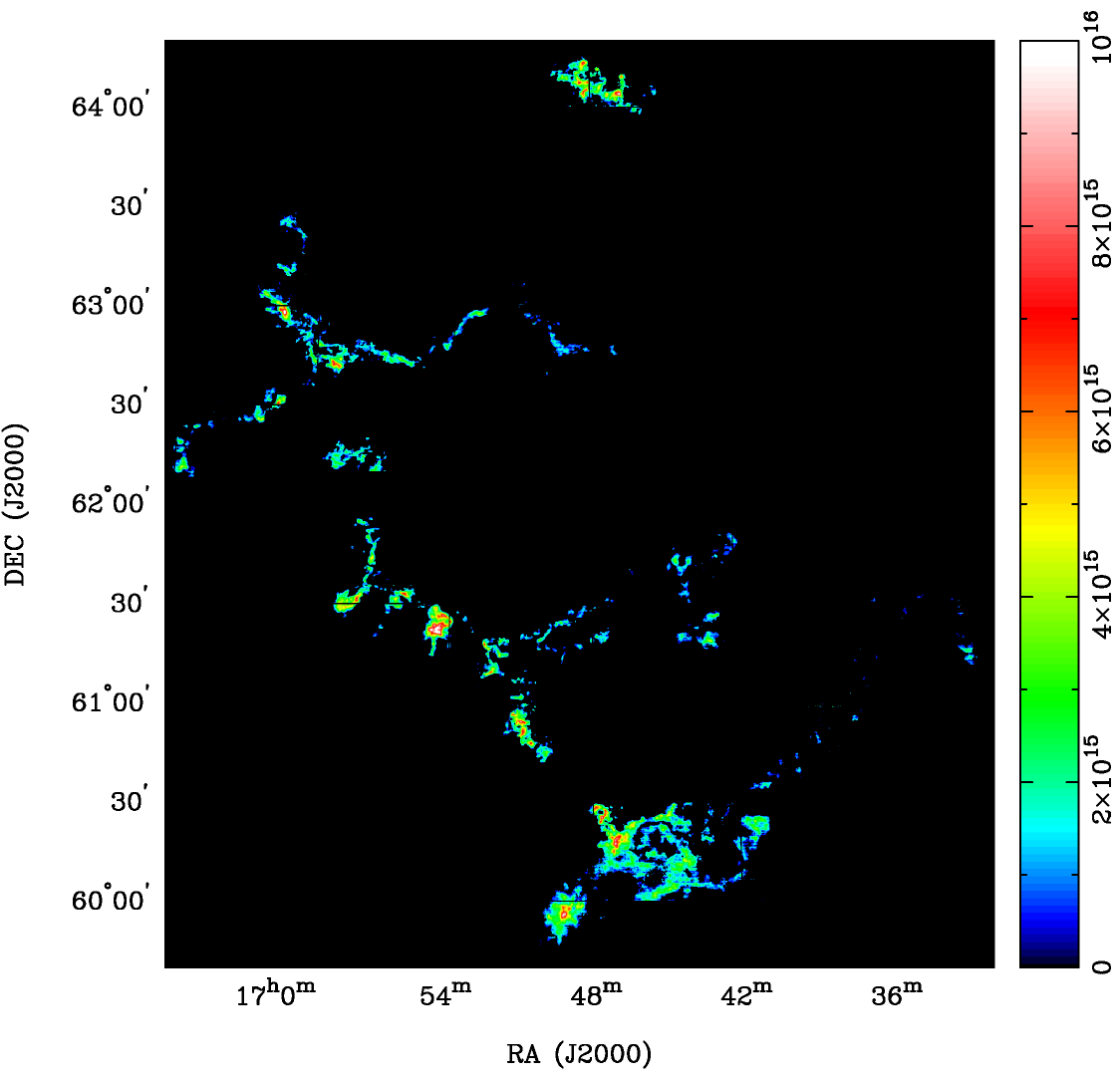}
    \includegraphics[width=3.22in]{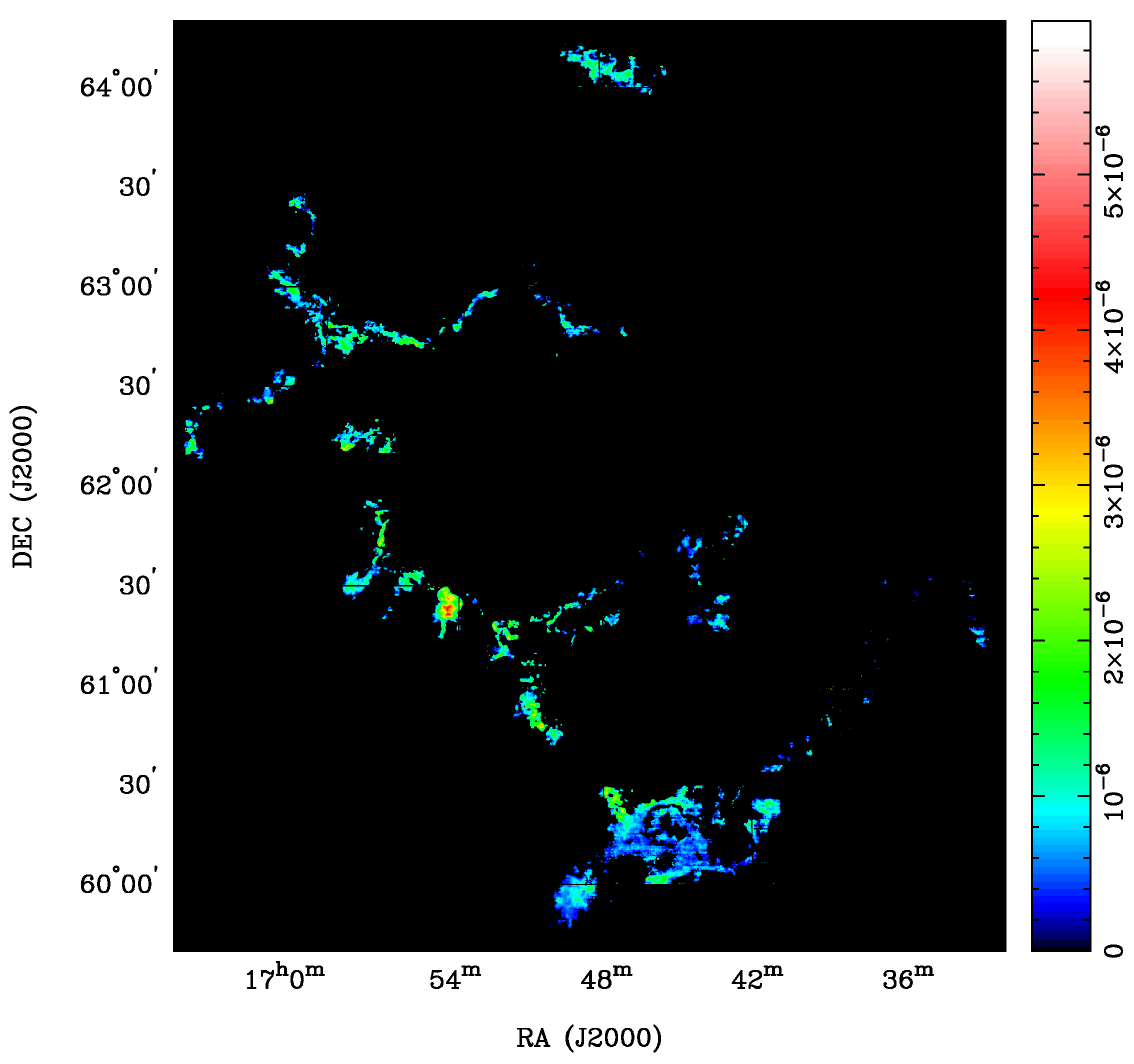}
    \caption{{\it (left:)} Total CO column density from LTE calculation at T$_{kin}$=20K.  Color wedge is labelled in CO molecules-cm$^{-2}$.  {\it (right:)}  Ratio of CO column density to total hydrogen column density (CO molecules/H-atom).  Resolution of both images is 38\arcsec.}
    \label{fig:ShuoTotalCOcolden}
\end{figure*}

The LTE map of total CO column density, N(CO), is shown in Figure \ref{fig:ShuoTotalCOcolden} {\it (left panel).}  There is considerable variation over the many discrete features detected in CO.  This variation is not simply a result of variation in the total gas density.   \citet{2017A&A...599A.109M} argue that the far-IR surface brightness from their {\it Herschel} 250 $\mu$m map is directly proportional to the total column density of gas, i.e., hydrogen in all forms.  We use their relation to derive a map of total hydrogen column density, N(H$_{total}$) and then divide our map of N(CO) by N(H$_{total}$) to find the distribution of CO abundance relative to total hydrogen.  The result is shown in Figure \ref{fig:ShuoTotalCOcolden} {\it (right panel).}  The relative abundance of CO varies by at least an order of magnitude over the Draco components where CO is detected.  One notable feature is the relatively high abundance peak seen in selected area C  (see Figure \ref{fig:Fig1_4combo}), where the ratio N(CO)/N(H$_{total}$) $\approx 5\times 10^{-6}$ (CO molecules/H atom).  More typically, this ratio is in the range $5\times 10^{-7}$~to~$1\times 10^{-6}$ over most of Draco.


\subsection{Derivation of molecular hydrogen column density and related quantities}

We want to compare the calculated CO column density with that of molecular hydrogen, of which CO is the most commonly used tracer.  The combination of the {\it Herschel} 250 $\mu$m dust continuum map and the atomic hydrogen maps from the DHIGLS survey can be used to infer the molecular hydrogen column density for the velocity range of the IVC, if we make certain reasonable assumptions.  \citet{2017A&A...599A.109M} argue that their far-IR continuum map is a good tracer for the total column density of hydrogen atoms in all forms over the mapped region and derive a simple linear relation between the 250 $\mu$m surface brightness, $I(250)$, and the column density of hydrogen in all forms, N(H$_{total}$).  We have opted to use their relation to derive a ``baseline" estimate of N(H$_{total}$).  (Two alternative analyses of the {\it Herschel} dust continuum maps by \citet{Singh_2022} and \citet{refId0} have reached somewhat disparate conclusions for the total dust and gas column density distributions, based on SED fitting of the far-IR data.  In Appendix \ref{Appendix}, we discuss the possible implications of those studies compared to our baseline assumption.)  For the present discussion, we use the simple relation $$N(H_{total}) = 2.49 \times 10^{20} I(250) {\rm ~atoms ~cm^{-2}}$$ from \citet{2017A&A...599A.109M}.

\begin{figure*}
    \includegraphics[width=3.2in]{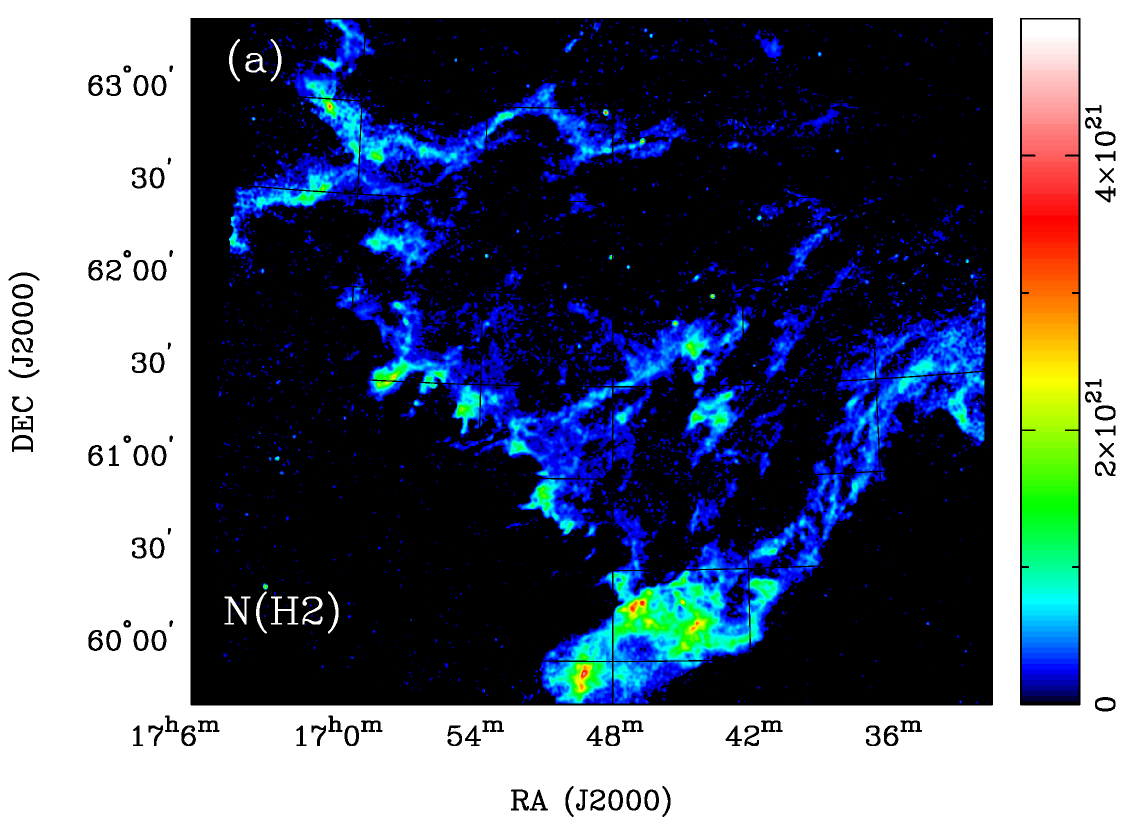}
    \includegraphics[width=3.2in]{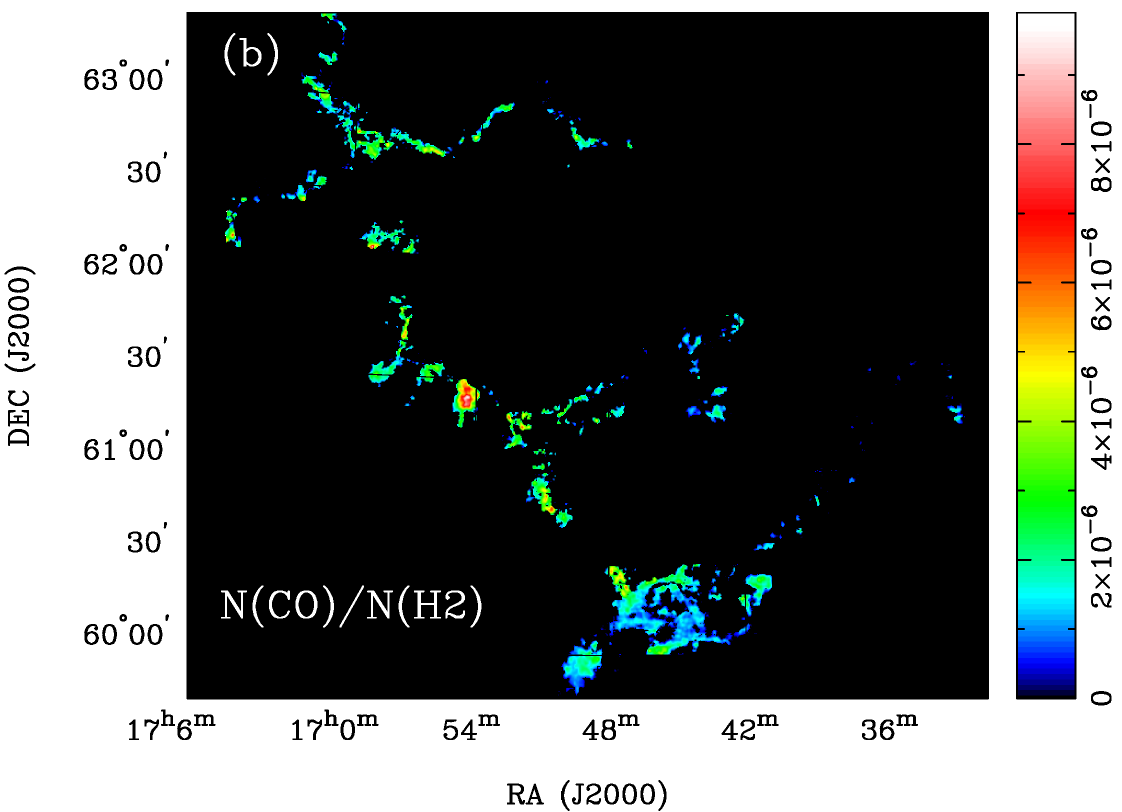}//
    \includegraphics[width=3.2in]{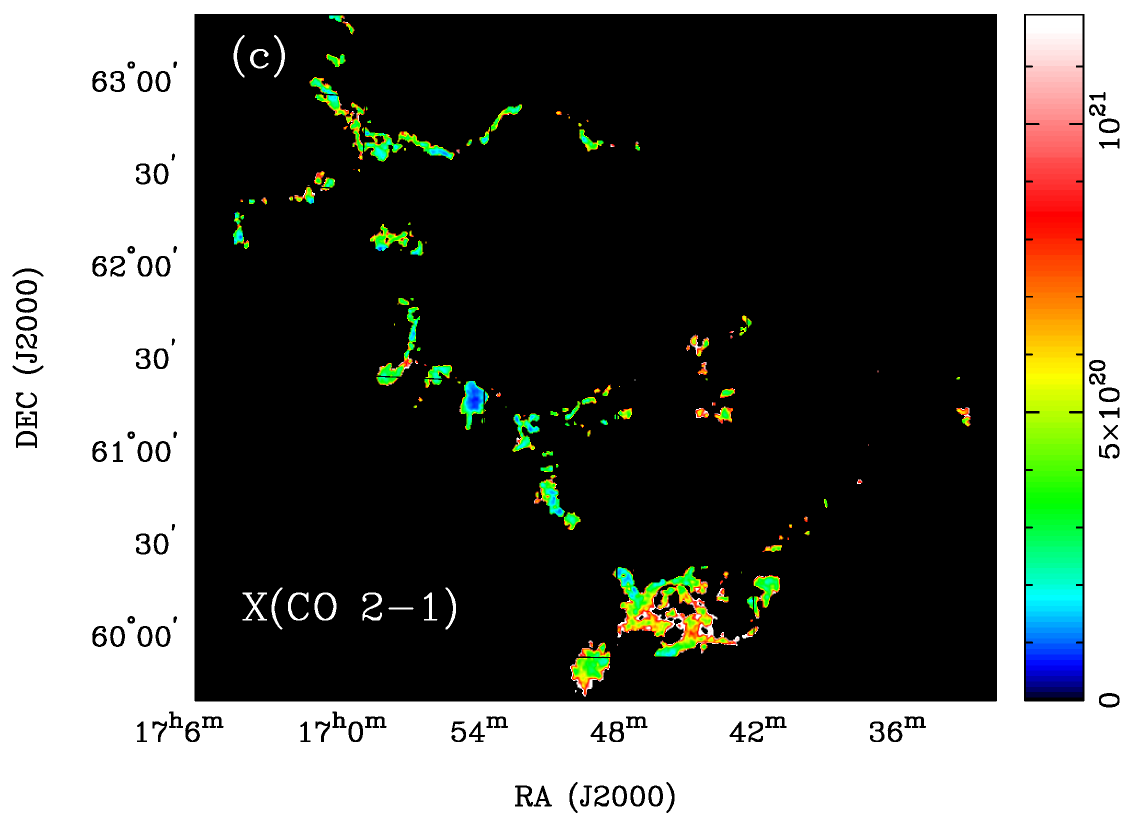}
    \includegraphics[width=3.2in]{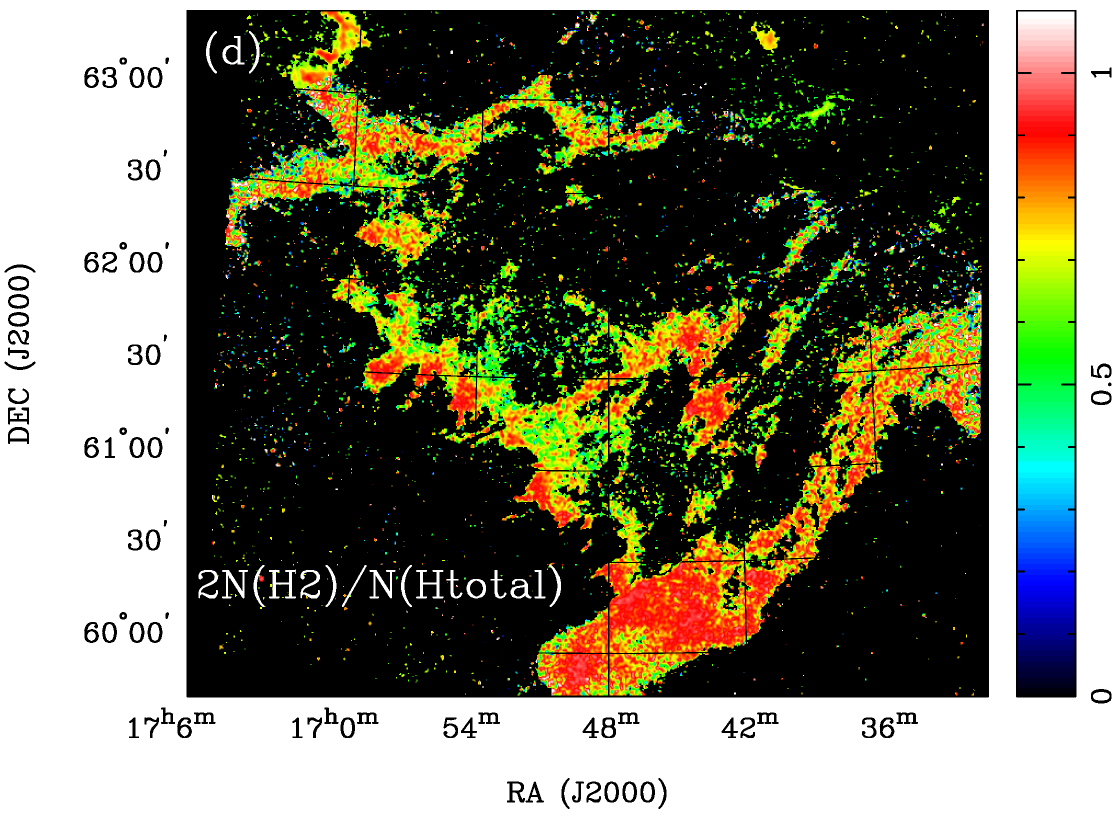}
    \caption{{\it (a)} H$_2$ column density in IVC velocity range.  The area displayed is limited to the DHIGLS field, from which the H$_2$ column density is derived by subtracting the HI column density from the {\it Herschel}-derived total hydrogen column density.  Color wedge is labelled in H$_2$ molecules cm$^{-2}$.  The resolution is 61\arcsec $\times$ 54\arcsec.  {\it (b)} Ratio of total column density of CO to column density of H$_2$.  {\it (c)} CO J=2-1 X-factor, in units of (H$_2$ molecules cm$^{-2}$)/(K-\kms).  {\it (d)}  Fraction of hydrogen atoms in molecular form in the DHIGLS field, i.e., 2N(H$_2$)/N(H$_{total}$).}
    \label{fig:H2_4panel_IVCmaps}
\end{figure*}

The DHIGLS map cube of atomic hydrogen from \citet{Blagrave_2017} can be divided into velocity ranges that cover the IVC and LVC emission separately.  Based on the lower resolution map of the GHIGLS survey, as shown in \citet{2017A&A...599A.109M}, we assume that the HI emission in the LVC can be neglected over the area of the dust structures in Draco.  Likewise, the HVC feature in the GHIGLS map largely avoids the Draco dust features, and since that range of velocities was not included in the DHIGLS map, we neglect the HVC as well.   Then after convolving the {\it Herschel} map to match the resolution of the DHIGLS image, which is 61\arcsec $\times$ 54\arcsec~ \citep{Blagrave_2017},  we subtract the HI column density for the IVC from the N(H$_{total}$) image.  The residual is assumed to represent the column density of molecular hydrogen, N(H$_2$).  This map is necessarily limited to the area covered in the DHIGLS image, which omits the top-most part of the area we mapped in CO.  The resulting map of molecular hydrogen column density in Draco is shown in Figure \ref{fig:H2_4panel_IVCmaps}(a).  

To compare the CO and H$_2$ column density maps, we smoothed and regridded the LTE-derived map of N(CO) (Figure \ref{fig:ShuoTotalCOcolden} left panel) to match the resolution and area of the map of N(H$_2$) over the DHIGLS field.  The ratio of these maps, N(CO)/N(H$_2$), is shown in Figure \ref{fig:H2_4panel_IVCmaps}(b).  The ratio varies by at least one order of magnitude over the map, from $\sim 1 \times 10^{-6}$ to $1 \times 10^{-5}$, with the most typical values around $3 \times 10^{-6}$.  The highest values are in selected field D (see Fig. \ref{fig:Fig1_4combo}), which stands out compared to the bulk of the molecular gas in having a relatively high CO abundance relative to H$_2$.  Even so, this peak abundance is about one order of magnitude lower than the typical CO/H$_2$ abundances inferred for most galactic Giant Molecular Clouds (GMCs).  In section \ref{Timescales} we discuss the implications of these CO/H$_2$ abundances and their spatial variations in terms of time-dependent models for molecule formation.

It is unlikely that the spatial variations in N(CO)/N(H$_2$) could result from small-scale spatial variations in the dust properties within and between clumps.  The maps of dust optical depth and dust temperature (\citet{Singh_2022}; P. Martin, private communication) show relatively small varations at the resolution of 36\arcsec.

Another quantity of interest is the so-called X-factor, which is sometimes used to estimate the total molecular column density, N(H$_2$), directly from the CO brightness temperature integrated over the line profile.  Most commonly this factor has been derived for the CO J=1-0 transition, simply because most observations of CO have previously been made of that line.  Here we calculate the corresponding quantity X(CO 2-1) as the ratio of the molecular hydrogen column density, N(H$_2$), divided by the integrated CO J=2-1 line intensity, I(CO 2-1).  We smoothed and regridded the integrated CO intensity shown in Figure \ref{fig:Fig1_4combo}(b) to match the image of N(H$_2$) shown in Figure \ref{fig:H2_4panel_IVCmaps}(a).  The ratio of the two images is defined as the X(CO 2-1) factor, i.e., X(CO 2-1) $\equiv$ N(H$_2$)/I(CO 2-1).  The resulting map is shown in Figure \ref{fig:H2_4panel_IVCmaps}(c).  It is evident that the X-factor varies by nearly an order of magnitude over Draco, from low values of order $1 \times 10^{20}$ to $1 \times 10^{21}$ [H$_2$ molecules cm$^{-2}$] [K-\kms]$^{-1}$.  The mean and median over the whole image are $5.2 \times 10^{20}$ and $4.5 \times 10^{20}$ [H$_2$ molecules cm$^{-2}$] [K-\kms]$^{-1}$ respectively, with a long tail to higher values.  \citet{Lewis_2021} found a comparably large variation in X(CO 2-1) for the California Molecular Cloud (CMC), where they derived total gas column densities from {\it Herschel} observations with SED fits to the far-IR emission.  They find values of X(CO 2-1) in a range from $(0.5 - 52) \times 10^{20}$ H$_2$ molecules cm$^{-2}$ [K-\kms]$^{-1}$, with an average over the whole CMC of $(4.1 \pm 0.8) \times 10^{20}$ H$_2$ molecules cm$^{-2}$ [K-\kms]$^{-1}$.  Their cloud-averaged mean is only about 20\% lower than the mean we derive for Draco.  For the CMC, however, they find a strong dependence of X(CO 2-1) on the dust temperature.  For sight-lines with T$_{dust} \geq 25$~K, they calculate a fairly small dispersion of values with an average X(CO 2-1) of $(1.3 \pm 0.4) \times 10^{20}$ H$_2$ molecules cm$^{-2}$ [K-\kms]$^{-1}$.  We point out that our selected region D has X(CO 2-1) close to the average of \citet{Lewis_2021} for the higher T$_{dust}$ pixels in the CMC.  Region D also shows the largest CO/H$_2$ abundance ratio in Draco.

The average abundance ratios N(CO)/N(H2) reflect the evolutionary state of the cloud, in terms of molecule formation.  The Draco molecular features must be very young compared to the CMC or other typical GMCs.  We find a lower average CO abundance N(CO)/N(H$_2$) for Draco compared to the CMC, even though the {\it averaged} X(CO)-factors are similar.  Time dependent numerical simulations, e.g. \citet{2018ApJ...858...16G} show, however, that the X(CO) factor depends on a variety of cloud properties as well as cloud age, including CO optical depth, gas temperature, incident radiation, cloud structure, etc.  The similarity of the {\it average} X-factors between Draco and the CMC is probably related to these other properties, and simply reinforces the conclusion that the CO X-factor is a poor measure of total gas mass on sub-parsec scales.

An earlier study of CO in parts of the Draco nebula by \citet{1993A&A...272..514H} used observations of the CO J=1-0 transition, together with dust continuum maps by {\it IRAS} and HI 21 cm maps with various radio telescopes, to determine total gas column densities and also the CO X-factor.  They derived X-factors that were an order of magnitude smaller than those we find here using the CO J=2-1 line.  This discrepancy is unlikely to be entirely a result of the different CO lines used.  \citet{1993A&A...272..514H} used the {\it IRAS} maps with angular resolution of 3\arcmin $\times$ 5\arcmin, but the {\it Herschel} 250 \um~images of \citet{2017A&A...599A.109M} show that the IVC features with detected CO emission are $\sim$1\arcmin~or less in size.  Consequently their measures of total column density for the IVC features are severely beam-diluted.  The telescopes they employed for CO mapping had resolutions of $\sim 4$\arcmin~ with spectra sampled on a 4\arcmin~ grid, so the effective resolution was $\sim 8$\arcmin.  We suspect that these differences in map resolution are the principal reason for the discrepancy in X(CO) from \citet{1993A&A...272..514H} compared with our values for X(CO 2-1). 

In the context of molecule formation, we want to know the fraction of total hydrogen that has been converted to molecular form.  We can only calculate a ratio of the column density-averaged fraction,  2N(H$_2$)/N(H$_{total}$) which is the line-of-sight averaged fraction of H atoms in molecular form.  The resulting map is shown in Figure \ref{fig:H2_4panel_IVCmaps}d.  The image shows that a significant fraction of the Draco cloud has been converted to molecular hydrogen especially in the ``leading edge" of the filamentary structures seen in the {\it Herschel} 250 $\mu$m map, and in the long concentration of emission on the southern part of the nebula.  In these regions the molecular fraction exceeds 90\%.  Other parts of the filaments have molecular fractions on the order of 50\% to 70\%.  Figure \ref{fig:H2_4panel_IVCmaps}d shows that in fact a significant portion of Draco is molecular, contrary to what is sometimes assumed in recent studies (e.g., \cite{refId0} ).

\section{Discussion}
\label{sec:Discussion}

\subsection{Uncertainties in derivation of column densities}
\label{Htotal}

Our analysis as summarized in Figure~\ref{fig:H2_4panel_IVCmaps} is based on 3 components of the gas content, all constrained to be the line-of-sight integrated column densities:  (1) the CO molecule, derived from our observations of the J=2-1 transition, where the total CO column density N(CO) is obtained from an assumed LTE level population at a kinetic temperature of 20 K;  (2) atomic hydrogen, where we rely on the DGHILS maps of \citet{Blagrave_2017} of the 21 cm HI line as a robust measure of N(HI);  (3) the total gas column density of hydrogen, N(H$_{total}$), which we obtain from the {\it Herschel} 250 micron intensity map of \citet{2017yCat..35990109M} and apply the simple proportionality of \citet{2017A&A...599A.109M}.  

If our assumption of LTE level populations for CO is correct, then the derived total CO column density is relatively insensitive to the assumed kinetic temperature.  Moreover, the lack of detected \tco~ and where detected, low brightness, confirms that the CO J=2-1 line is generally optically thin across the nebula.  We assume that T$_{kin} = 20$~K, but for values in the range $14 - 25$~K the LTE total column density varies by only a few per cent.  Since the J=2 level lies at $\sim 16$~K above ground, the levels J=1 and 2 should be nearly thermalized provided the gas density is high enough.  If observations of the J=1-0 and 3-2 lines with sensitivity and resolution comparable to our maps were available, a comparison could confirm whether our assumption of LTE is correct.  Absent such data, we regard our approach as the least subject to other uncertain quantities that would affect, for example, a Large Velocity Gradient analysis.

The column density of atomic hydrogen is very well determined by the integrated intensity of the 21 cm HI line, since the emission is nearly independent of the gas temperature.  We take the DHIGLS map of the IVC gas from \citet{Blagrave_2017} as a firm lower limit on the total gas column density.  The derived N(HI) values are somewhat compromised, however, by the modest resolution ($\sim 1\arcmin$) and more seriously, the limited sensitivity of their data.  Even so, the HI map of the IVC emission is a critical constraint on the total gas column density for the area of Draco covered by the DHIGLS field.

We therefore argue that the maps of both N(CO) and N(HI) are reasonably secure.  A more serious concern is the determination of the total gas column density as inferred from various analyses of the far-IR thermal dust emission measured by the {\it Herschel} PACS and SPIRE cameras.  In this paper, we will adopt as a ``baseline" the simple proportionality from \citet{2017A&A...599A.109M}, namely that N(H$_{total}) = 2.49 \times 10^{20}~ I(250)$ H-atoms cm$^{-2}$ where $I(250)$ is the zero-corrected intensity in MJy/sterad at wavelength 250 $\mu$m from the SPIRE instrument.  
Since the publication of \citet{2017A&A...599A.109M} there have been two other analyses of the {\it Herschel} maps of Draco which attempt a more refined treatment of the thermal dust emission and from that, to infer the distribution of total gas column density.  In Appendix A, we consider the implications of these studies with respect to our ``baseline" map described in Section \ref{sec:LTEanalysis}.  In the following Discussion, however, we will apply the simple proportionality of \citet{2017A&A...599A.109M} for the calculation of total hydrogen column density.

\subsection{Timescales for molecule formation}
\label{Timescales}

A significant objective of our Draco observations is to provide constraints on models of molecule formation in interstellar clouds.  Constructing reasonably plausible simulations of this process is an exceedingly complex challenge for numerical computations, so most published results to date are restricted to simplified versions of either cloud structure and dynamics and/or of relevant chemical processes (typically including ion-molecule, neutral radical, and grain surface reactions, as well as photo-processes).  Well-resolved time-dependent models in particular are rare.  Moreover, the Draco cloud is substantially different from typical molecular clouds in the Galactic disk, as we noted above.  Despite these caveats, the properties derived from our CO observations and related published work offer some useful data applicable to numerical simulations of molecule formation.  

The first constraint we found is that CO only becomes detectable for a {\it Herschel} I(250\um) $\ge 5$~MJy/sr, which corresponds to a minimum total column density of H in all forms of $\sim1.3\times 10^{21}$~ H-atoms/cm$^2$ for the onset of CO formation, if we assume the conversion factor in \citet{2017A&A...599A.109M}.  Above this intensity threshold, there is only a fair correlation between I(250\um) and the integrated CO J=2-1 line intensity when plotted over all the detected pixels in the entire field mapped in CO.  As shown in Figures \ref{fig:BoxAchanmaps} - \ref{fig:BoxEchanmaps}, however, the remarkably complex velocity structure of the molecular gas is not revealed in the far-IR continuum.  The inability of using the thermal dust continuum to determine total gas column density as a function of radial velocity is perhaps the most important limitation on deriving relative molecular abundance distributions, e.g., CO/H$_2$, for comparison with numerical models.  The substantial uncertainties in the assumed gas/dust mass ratio and the dust mass opacity coefficient appropriate for Draco add additional potential sources of error in quantitative comparisons.

\begin{figure*}
    \includegraphics[width=6.in]{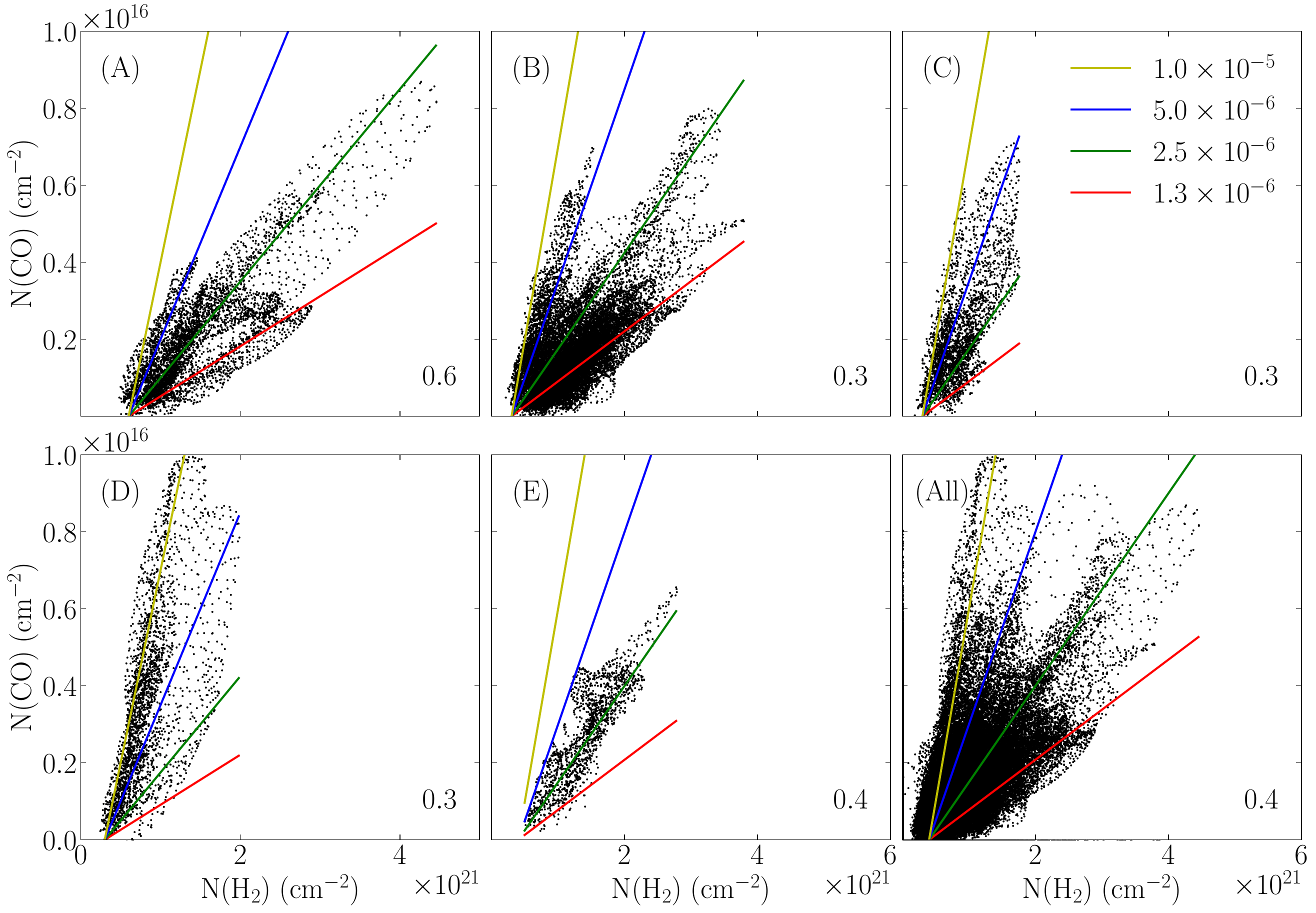}
    \caption{Scatter plots of individual (10\arcsec)$^2$ pixel values of N(CO) vs. N(H$_2$) (in molecules/cm$^2$) for selected regions A - E and for the combined set.  Colored lines have slopes as labelled, which give the asymptotic value for the ratio N(CO)/N(H$_2$). The x-intercept values in 10$^{21}$ H$_2$-molecules cm$^{-2}$ is given at lower right corners in each panel. }
    \label{fig:H2_CO scatterplots}
\end{figure*}

Despite these limitations, we found that correlation scatter diagrams of our derived column densities of CO vs. molecular hydrogen show clear trends, when restricted to the smaller regions A - E shown in Figure \ref{fig:Fig1_4combo}.  These comparisons are illustrated in Figure \ref{fig:H2_CO scatterplots}.  In each region, there are one or more distinct ``streams" of points which are nearly linear with different slopes but intersecting the x-axis at about the same value of N(H$_2$).  The colored lines show 4 examples with slopes as labelled.  In each panel, all 4 lines have the same x-intercept, which differ from panel to panel but have a mean of  N(H$_2$) $= 4 \pm 1 \times 10^{20}$ H$_2$ molecules cm$^{-2}$.   (These examples are by-eye estimates, not linear regressions, given the large scatter.)   The x-intercepts imply that the CO J=2-1 line is detected only where the column density of molecular hydrogen corresponds to a {\it Herschel} I(250) $\ge$ 3 MJy/sr, assuming all hydrogen is molecular and using the scale factor from \citet{2017A&A...599A.109M}.  This lower limit is consistent with the value noted above of 5 MJy/sr if, as is likely, some hydrogen is in atomic form at the outer edges of the molecular gas.  

The approximately linear appearance of the ``streams" in Figure \ref{fig:H2_CO scatterplots} implies also that the abundance ratio of column densities, N(CO)/N(H$_2$) rises from zero at the x-intercept value to an asymptotic limit given by the slope of each line.   It is striking that three of the 5 selected fields, namely C, D, and E, appear to have one dominant feature, but with clearly different slopes between the 3 fields, while two fields, A and B, show two or three distinct ``streams" each with different slopes.   Fields A and B are spatially larger and have more structure in velocity than the other three fields, suggesting that the multiple streams of points are associated with distinct spatial/kinematic components.  The differences in slopes imply differences in the current molecular abundance ratio of CO/H$_2$ between and within the 5 selected fields.  The high-resolution time-dependent chemical modelling of \citet{10.1111/j.1365-2966.2009.15718.x} suggests that the explanation for the different slopes lies in the time since molecule formation began in each feature.  Their chemical evolution models for formation of CO (see their Figure 5) show a steep initial rise in the mass-weighted CO/H$_2$ abundance, increasing from $10^{-6}$ to $10^{-5}$ in only about 300,000 years.   

For comparison, we can estimate the dynamical timescales for the individual features in Draco from our measured CO radial velocities as shown in Figure \ref{fig:Fig1_4combo}.  A typical CO line has a radial velocity of $-20$ \kms.  If we assume that the spatial motion is directly toward the Galactic plane, at the latitude of 38\degree ~the total spatial velocity is $-33$ \kms~ and the transverse velocity is $26$ \kms.  At our adopted distance of 600 pc, the transverse motion of a feature would be about 15\arcmin ~in $10^5$ yr.  This angular size is comparable to the dimensions of the selected fields illustrated above, for example field D (Figure \ref{fig:BoxDchanmaps}).  The multiple streams of points seen in fields A and B (Figure \ref{fig:H2_CO scatterplots}) may indicate that in these spatially larger fields, individual clumps of gas have undergone formation of CO with different timescales which correspond to the dynamical timescale for the infalling gas.  In this picture, the streams with the steepest slopes (e.g, field D) have been forming CO for longer times than those with shallower slopes (e.g, field E).  Since we do not know the detailed history of the gas motions in Draco, we cannot make specific conclusions about molecular (i.e., CO) formation in the individual features in our maps.  Our observations do appear to be consistent with the time-dependent models of \citet{10.1111/j.1365-2966.2009.15718.x}, who find that molecule formation is rapid once gas densities reach high enough values, and that the CO/H$_2$ abundance varies by large factors within a clumpy turbulent cloud.  

\subsection{Dynamics}
\label{Dynamics}
An important motivation for our spectroscopic mapping is the capability to distinguish kinematic features in the radial velocity dimension, given the very high spectral resolution factor $R = \nu_o/\delta\nu = (230 \rm{~GHz}/250 \rm{~kHz}) \approx 10^6$ of the data.  This capability is shown in Figures \ref{fig:BoxAchanmaps} - \ref{fig:BoxEchanmaps}, which demonstrate that the structures found in the {\it Herschel} far-IR continuum maps typically contain individual molecular components with relatively narrow radial velocity widths, but which span a much broader range of velocities over the entire structure.  These dynamical properties are also evident in the corresponding spectra shown in Figure \ref{fig:SpecPlotsSelectedRegions}.  The first moment or velocity centroid maps of the five selected fields A - E span between 4 and 8 \kms, while the corresponding line-widths (the second moment or velocity dispersion) at each pixel typically show a single peak at 0.5 \kms ~and extend to less than 2 \kms.  This kinematic behavior is seen over the entire region mapped in CO, as shown in 
 the lower two panels of Figure \ref{fig:Fig1_4combo}.  The many individual clumps ($\sim 5000$) described by \citet{2017A&A...599A.109M} from their 250\um ~image evidently have internal velocity motions close to the thermal width for gas temperatures of order 20 K, but these clumps have significantly larger (i.e., supersonic) systemic velocities relative to each other.  This important kinematic information cannot be extracted from the thermal dust continuum emission, but is clearly a critical constraint on numerical simulations of the dynamics of the Draco cloud.  

The narrow CO line-widths of individual gas clumps do appear to be consistent with the estimates by \citet{2017A&A...599A.109M} for the viscosity and corresponding turbulent energy dissipation scales in Draco.  They found that kinematic viscosity must be dominant compared to molecular viscosity, and from that they estimate a typical length of 0.1 pc for dissipation of supersonic turbulence.  (They also estimate a comparable scale length of 0.1 pc for energy dissipation by ion-neutral friction, i.e., ambipolar diffusion, assuming reasonable values for the gas density and magnetic field strength.)  The infalling IVC cloud has a Mach number of $\sim 6$, according to \citet{2017A&A...599A.109M}.  This value is comparable to the spread of velocities in the clumpy structures of the individual R-T features highlighted in selected fields A - E (Figures \ref{fig:BoxAchanmaps} - \ref{fig:BoxEchanmaps}, also Figure \ref{fig:SpecPlotsSelectedRegions}).   We observe that the individual molecular clumps have CO line-widths close to thermal or slightly trans-sonic, where the spatial resolution of our CO map is 0.11 pc.  If the turbulent dissipation scale were much less than 0.1 pc, we would expect to find CO line-widths well above the thermal value, since each resolution element in our maps should encompass a range of turbulent velocities along the line of sight.  In this picture, the velocity structure of the R-T features was determined by supersonic turbulence as the IVC encountered the WNM layer, while the internal velocity spread of the individual molecular clumps has been reduced to near-thermal widths by energy dissipation mediated via kinematic viscosity and possibly ambipolar diffusion.  If the scale length for dissipation of turbulence is in fact about 0.1 pc we would predict that CO J=2-1 observations with higher angular resolution than ours, e.g. $\sim$10\arcsec, would reveal some clumps with even narrower line-widths.  The time-scale for turbulence dissipation within a molecular clump is evidently comparable to the chemical formation time-scale, i.e. on the order of a few times $10^5$ years, but with some variation between clumps as implied by the variation in slopes seen in Figure \ref{fig:H2_CO scatterplots}.

We note that a numerical simulation by \citet{2014A&A...567A..16S} showed that compressive turbulence in the WNM can lead to rapid cooling and formation of dense structures within the WNM, with time-scales $< 10^6$ yr.  This simulation was not precisely comparable to the scenario proposed to explain the morphology of Draco, but does suggest that supersonic turbulence created as the infalling IVC encounters the WNM layer in the Galactic plane can indeed lead to the conditions needed for rapid formation of CO molecules as found by \citet{10.1111/j.1365-2966.2009.15718.x}.  Our observations of CO emission could provide constraints on numerical simulations which combine the dynamics of the formation of R-T structures by infalling gas, together with the chemistry of H$_2$ and CO.  

\subsection{Future evolution of Draco}
\label{Draco future}
We can speculate about the future evolution of the Draco IVC based on the available observations and analysis.  If the nebula is now 370 pc above the Galactic plane and falling in with a velocity of $\sim 40$~\kms, and continues to move at that speed, it will take about 10 million years to reach the mid-plane.  If the gas is slowed as it encounters more of the WNM, as seems likely, the time to reach the minimum in the stellar gravitational potential would be longer.  The denser clumps we see in the R-T structures may be less affected by such slowing than the more diffuse parts of the IVC, which would lead to some vertical stretching or separation of the IVC material.  The denser parts might even overshoot somewhat before being drawn back by the gravity of the stellar disk, in a damped oscillatory motion, eventually settling into the galactic mid-plane in a few tens of millions of years, i.e, short compared to the galactic orbital timescale for the solar neighborhood.

\citet{2017A&A...599A.109M} used the {\it clumpfind} algorithm to identify about 5000 distinct condensations in their {\it Herschel} 250 \um ~map.  They calculated a total mass for these condensations of $\sim 5000$ \msun, assuming the standard gas/dust mass ratio and dust opacity as discussed in Section \ref{Htotal} above.  They also derived masses for the individual clumps in the range 0.1 to 20 \msun, with a median mass of 0.53 \msun.  The great majority of these condensations were not gravitationally bound, however, with only 15\% of the total mass in bound structures.  
Provided these clumps remain together as an identifiable assembly, the future appearance of Draco may be something like the dark clouds we see in the solar neighborhood, at the lower end of the mass spectrum of galactic molecular clouds, e.g. Table 32.2 in  \citet{Draine2011}.  The bulk of the infalling gas will likely disperse into the galactic interstellar medium on a timescale of a few times 10$^7$ years.  The surviving more massive molecular condensations may eventually form low mass stars in a sparse unbound stellar cluster.

\section{Summary}

We present new observations made with the Steward Observatory Heinrich Hertz Submillimeter Telescope, of the J=2-1 emission from CO toward the Draco Nebula Intermediate Velocity Cloud (IVC).  The map covers about 8500 square arcminutes, with an angular resolution of 38\arcsec~ and spectral resolution of 0.33 \kms~ over the velocity range of the IVC.  The area mapped was selected to cover all of the thermal dust emission seen in the 250\um~ {\it Herschel} image published by \citet{2017A&A...599A.109M} with intensity I(250) $\geq$ 5 MJy/sr, which marks the lower limit at which CO is detected.  From the relationship in \citet{2017A&A...599A.109M} between I(250) and total hydrogen column density, N(H$_{total}$) as a ``baseline", we derive a map of molecular hydrogen column density, N(H$_2$) by subtracting from N(H$_{total}$) the atomic hydrogen column density in the IVC mapped in the DHIGLS survey \citep{Blagrave_2017}.  The column density of CO molecules is obtained from an LTE analysis, under the assumption that the CO J=2-1 line is optically thin over the nebula.  This assumption is justified by the observed lack of \tco~ emission except in a few positions.  We emphasize also that the derived total CO column density, N(CO), is insensitive to the assumed gas kinetic temperature for the plausible range of values.  From these column density maps, we present maps of the molecular abundance ratio N(CO)/N(H$_2$).  

In the Appendix, we also discuss two recent alternative analyses \citep{refId0,Singh_2022} of the thermal dust continuum maps by the {\it Herschel} and {\it Planck} satellites, in terms of the derived total gas column density.  To reconcile these analyses, we find that our ``baseline" N(H$_{total}$) map would have to be scaled down by a factor of 0.3, assuming that their adopted gas-to-dust ratios and/or dust mass opacity coefficients are not correct for Draco compared with matter in the Galactic plane.  

The Draco IVC shows a remarkably clumpy structure, with ``fingers" of dust emission extending toward the Galactic plane.  These features have been interpreted by \citet{2017A&A...599A.109M} as arising from a Rayleigh-Taylor (R-T) instability that develops when the in-falling atomic gas in the IVC encounters the Warm Neutral Medium at supersonic velocities.  The high resolution in velocity of our CO map spectra allows a detailed examination of the kinematics of this process, and of the formation of CO in the in-falling gas.  Our main conclusions are:

(1)  CO only becomes detectable for a {\it Herschel} I(250\um) $\ge 5$~MJy/sr, which corresponds to a minimum total column density of H in all forms of $\sim1.3\times 10^{21}$~ H-atoms/cm$^2$ for the onset of CO formation, if we assume the conversion factor in \citet{2017A&A...599A.109M}.   The derived CO column density does not depend on the total hydrogen column density, so if we scale down N(H$_{total}$) by a factor of 0.3 as suggested in the Appendix, the onset of CO formation would be where N(H$_{total}$) exceeds about $4\times 10^{20}$~H-atoms/cm$^2$.

(2)  The ratio of CO to molecular hydrogen column densities, N(CO)/N(H$_2$), varies by at least an order of magnitude across the nebula, ranging from $10^{-6}$ to $10^{-5}$ as shown in Figure \ref{fig:H2_4panel_IVCmaps}b.  Even the highest ratio is about a factor of 10 less than is typically found in the Galactic GMCs.  The fraction of hydrogen in molecular form (Figure \ref{fig:H2_4panel_IVCmaps}d) shows a gradient along the R-T ``fingers" with values of $\sim 90\%$ at the highest toward the SE and falling to $\sim 30\%$ away from the peaks.  The large southern concentration and the sinuous filament in the northern part of Draco, however, have a high molecular fraction over most of their extent.  

(3)  Correlation diagrams of individual pixel values for N(CO) vs. N(H$_2$), when plotted for isolated map fields as in Figure \ref{fig:H2_CO scatterplots}, show well-defined linear features (``streams") with a range of slopes but similar x-axis intercepts.  We interpret these streams as indicators of CO formation that is occurring in the dense gas clumps but with a spread in timescales since the onset of chemical evolution.  The inferred abundance ratios, N(CO)/N(H$_2$), are consistent with the evolutionary models of \citet{10.1111/j.1365-2966.2009.15718.x} with ages spanning a few times 10$^5$ years.  

(4)  The very high spectral resolution of our data ($R \approx 10^6$) allows a detailed kinematic analysis of the CO emitting gas.  Expressed as velocity moments (centroid and dispersion) we find that the spread in centroid velocities is large compared to the dispersion of spectra at individual map pixels.  The spread in centroids over fields of $\sim15$\arcmin~ in size is comparable to the predicted sonic Mach number of the R-T structures \citep{2017A&A...599A.109M} with $M_{sonic}\approx$~6.  The velocity dispersions show a peak at $\sim$0.5~\kms, which is near the thermal line width for a kinetic temperature of 20 K, or at most is slightly trans-sonic.  We argue that these values for the distribution of velocity centroids and dispersions are consistent with turbulent motions driven as the IVC gas falls in and encounters the WNM.  The scale length for dissipation of turbulence is $\sim$0.1 pc, as predicted by \citet{2017A&A...599A.109M} in their model for formation of R-T instabilities.  The energy dissipation must then be mediated by kinematic viscosity and/or ambipolar diffusion, rather than molecular viscosity. 

(5)  The Draco nebula should settle onto the Galactic mid-plane in $\sim$$10^7$~years;  it will then resemble a dark cloud such as those seen in the Solar neighborhood.  The densest clumps may eventually form low mass stars in an unbound cluster, while the bulk of the gas will merge with the general interstellar medium.  

\section*{Acknowledgements}

The Heinrich Hertz Submillimeter Telescope is operated by the Arizona Radio Observatory, which is a unit of Steward Observatory at the University of Arizona.  We thank P.G. Martin, M.-A. Miville-Deschenes, V. Ossenkopf-Okada, and N. Schneider for useful comments and for providing maps in digital form.  
%%%%%%%%%%%%%%%%%%%%%%%%%%%%%%%%%%%%%%%%%%%%%%%%%%
\section*{Data Availability}

The {\it Herschel} 250 $\mu$m map used in this study is available at \citet{2017yCat..35990109M}.  The 21 cm HI data from the DHIGLS survey \citep{Blagrave_2017} are available at the website https://www.cita.utoronto.ca/DHIGLS/.  The CO J=2-1 map cube is available in the Harvard Dataverse (https://dataverse.harvard.edu) with DOI https://doi.org/10.7910/DVN/7DO6EP.



\bibliographystyle{mnras}
\bibliography{references}

\appendix
\section{Comments on alternative analyses of thermal dust continuum}
\label{Appendix}

Since the publication of \citet{2017A&A...599A.109M} there have been two other analyses of the {\it Herschel} maps of Draco which attempt a more refined treatment of the thermal dust emission and from that, to infer the distribution of total gas column density.  Here we consider the implications of these studies as regards our ``baseline" map described in Section \ref{Published observations}.  

\citet{Singh_2022} made an extensive analysis of {\it Herschel} PACS/SPIRE maps of a number of molecular clouds to derive the distributions of dust temperature and optical depth in their HOTT ({\it Herschel Optimized Tau and Temperature}) survey, with data available at www.cita.utoronto.ca/HOTT.  They made modified blackbody fits using the standard parameterization
$$I_{\nu} = \tau_{\nu_o} B_{\nu}(T_{dust}) {(\frac{\nu} {\nu_o})}^\beta$$
where the dust optical depth at a fiducial frequency, $\tau_{\nu_o}$, is given by
$$\tau_{\nu_o} = \kappa(\nu_{o}) r \mu m_H N(H). $$  
Here $r$ is the dust-to-gas ratio by mass, and $N(H)$ is the column density of hydrogen atoms in all forms, $\mu$~is the mean AMU mass per gas particle, and $m_H$~is the mass of the hydrogen atom.  The dust mass opacity coefficient, $\kappa(\nu)$ is assumed to have a power-law frequency dependence. \citet{Singh_2022} find the power-law exponent $\beta$~ by interpolating the lower resolution maps of $\beta$ from the Planck XI dust model \citep{2014A&A...571A..11P}.  Free parameters that were fitted to the intensity $I_\nu$ in the 160\um~ PACS and 250, 350, and 500\um~ SPIRE images at each pixel were the Planck dust temperature, $T_{dust}$ and the dust optical depth. %The dust optical depth $\tau_{\nu_o}$ is referred to a fiducial frequency, $\nu_o$. 
Though their published results concentrated on the clouds in the Gould's Belt Survey \citep{2010A&A...518L.102A} and did not explicitly include the Draco field, that region was also analyzed and kindly provided to us by P.G. Martin (private communication), in the form of maps of $\tau_{1~\rm{THz}}$~and $T_{dust}$, as well as a map of total hydrogen column density, N(H$_{total}$), which assumes nominal values of the dust/gas mass ratio $r = 1/100$, and the dust mass-opacity coefficient, $\kappa_{1~\rm{THz}} = 10$~ cm$^2$ g$^{-1}$.  We convolved and re-gridded their map of N(H$_{total}$) to match the map of HI in the IVC from the DHIGLS survey, and examined the ratio of the two images.  If the DHIGLS map is a firm lower limit to the total gas column density, as we assume, then the ratio N(H$_{total}$)/N(HI) must be greater than 1 everywhere.  Applying the nominal values for gas/dust mass ratio and $\kappa_{1~\rm{THz}}$ we find that many pixels have a ratio <1, implying that either one or both of the nominal values are incorrect for the Draco dust and gas.  \citet{Singh_2022} found that the dust temperatures in their modified black-body fits were typically in the range 19 - 20 K and ranged down to 17 K for the highest opacity pixels.  Thus, even at the higher angular resolution of {\it Herschel} compared to {\it Planck}, they found that the dust temperature is relatively uniform over the Draco field.

The DHIGLS N(HI) map does have a significant amount of scatter as a result of noise.  We found that, after regridding the \citet{Singh_2022} map of N(H$_{total}$) to match the DHIGLS N(HI) map, a plot of individual pixel values of the ratio N(H$_{total}$)/N(HI) on the y-axis, versus the integrated CO J=2-1 line intensity (a directly observed quantity) showed a clear trend such that the ratio N(H$_{total}$)/N(HI) increases with CO line intensity consistent with the assumption that CO traces H$_2$.  Given the condition that N(H$_{total}$)/N(HI) $\geq 1$, the trend line must intersect the y-axis at a value of at least 1, where all the gas is atomic.  Values greater than 1 imply a fraction of the gas has become molecular, as expected by the detection of CO.  For the selected regions, we found that the trend lines intersect the (ratio) y-axis at values in the range 0.3-0.7, i.e, well below the expected minimum value of 1.  This discrepancy suggests that the nominal values assumed for $\kappa_{\nu_o}$ or the dust/gas mass ratio or both, are not correct for the dust and gas in Draco.  The nominal values used here are likely more appropriate for molecular clouds in the Galactic plane.  If the matter in Draco was ejected from the plane by some combination of stellar winds and supernovae and is now falling back, i.e., the ``Galactic fountain" scenario, it is likely that the dust has been altered in the process to have properties rather different from its original state.  If icy mantles were lost, for example, the dust mass absorption coefficient $\kappa_{\nu_o}$ may be changed from the assumed 10 cm$^2$ g$^{-1}$.  If some fraction of the dust mass were destroyed then the assumed dust/gas mass ratio of 1/100 may well be too high  \citep{1979ApJ...231..438D,1983ApJ...275..652S,1996ApJ...469..740J,2004ApJ...614..796S}.  (In a recent study of 94 high latitude QSOs, \citet{Shull_2024} find that the average dust/gas ratio is a factor of $\sim 1.5$ lower than the standard value for gas in the Galactic plane, with some sight-lines showing values substantially lower than the average.)  A combination of such effects could easily combine to make an increase by the factor of $\sim 3$ necessary to bring the HOTT N(H$_{total}$) into agreement with the minimum value set by the DHIGLS N(HI) map for the IVC.  

A study by \citet{refId0} also analyzed the same {\it Herschel} observations of the Draco field and reached somewhat different conclusions.  They made fits to the SEDs using the 160\um~ PACS and 250, 350, and 500\um~SPIRE  images, all convolved to the lowest resolution (36.7\arcsec).  In contrast to \citet{Singh_2022}, \citet{refId0} fixed the frequency-dependent power law index $\beta$ at 2.0 and set the dust mass opacity coefficient to $\kappa_{\nu} = (0.13/r) ({\nu}/{1.2 {\rm~THz}})^\beta$~cm$^2$ g$^{-1}$, following \citet{2015A&A...584A..93J} for $\lambda = 250$\um ~($\nu = 1.2$~THz).   \citet{refId0} then fitted as free parameters the dust temperature and column density.  If we assume a dust/gas ratio of $r = 1/100$, as in the HOTT column density map from P. Martin (p.c.), the value of the dust mass opacity coefficient scaled to 1 THz, $\kappa$(1 THz) is 9.0 cm$^2$~g$^{-1}$ for \citet{refId0}, slightly less than 10 cm$^2$ g$^{-1}$ from \citet{Singh_2022}, and the assumed frequency dependence is steeper than found by \citet{Singh_2022}.   \citet{refId0} also find significantly lower dust temperatures compared to \citet{Singh_2022} with values distributed mainly in the range 11 - 13 K.  (Digital versions of the column density and dust temperature maps from \citet{refId0} were kindly provided to us by N. Schneider and V. Ossenkopf-Okada.)

The main difference between these two analyses is evidently the difference in adopted values for $\beta$.  \citet{refId0} assumed a fixed value of $\beta = 2$ while \citet{Singh_2022} use values which range between 1.4 and 1.7 with a mean of about 1.55, by interpolating from the sensitive but lower resolution maps of the {\it Planck} collaboration.  Since the assumed dust mass opacity coefficients are not very different ($\sim 10$\%) at the fiducial frequency of 1 THz, it is the difference in the adopted frequency dependencies that leads to different fits to the SED for dust temperature and optical depth.  Lower opacity coefficients coupled with lower fitted dust temperatures ($\sim12$~K vs. $\sim19$~K), hence lower values of $B_\nu(T_{dust})$, require greater dust column densities to produce the observed far-IR intensities.  As a consequence, \citet{refId0} find higher total column densities than \citet{Singh_2022} for the same assumed gas/dust mass ratios.  (See also Appendix C in \cite{Shull_2024} for further discussion of the uncertainties in dust models applied to far-IR images.)

In conclusion, we have opted to use the map of total hydrogen column density derived from the analysis of \citet{2017A&A...599A.109M} in our comparisons with the CO and HI distributions.  The alternative approaches of \citet{Singh_2022} and of \citet{refId0} have sufficiently large uncertainties in the critical dust-related parameters that we prefer the analysis rooted in the well-calibrated images from the {\it Planck} Collaboration as applied by \citet{2017A&A...599A.109M} to their 250 $\mu$m image. 

%%%%%%%%%%%%%%%%%%%%%%%%%%%%%%%%%%%%%%%%%%%%%%%%%

%%%%%%%%%%%%%%%%%%%%%%%%%%%%%%%%%%%%%%%%%%%%%%%%%%

% Don't change these lines
\bsp	% typesetting comment
\label{lastpage}
\end{document}